\documentclass[11pt,preprint]{aastex}
\usepackage{url}
\usepackage[ps2pdf]{thumbpdf} 
\usepackage[bookmarks]{hyperref}
\usepackage{natbib}

\newcommand{\HII}{H {\sc ii}}
\newcommand{\lsun}{\ensuremath{\mathrm{L}_\odot}}
\newcommand{\msun}{\ensuremath{\mathrm{M}_\odot}}
\newcommand{\ha}{H$\alpha$}
\newcommand{\halpha}{H$\alpha$}
\newcommand{\ang}{\AA}
\newcommand{\mult}{$\times$}
\newcommand{\av}{$A_{\rm v}$}

\newcommand{\mstar}{\mbox{$M_{\ast}$}}

\slugcomment{Draft v3}
\shorttitle{Triggered star formation in SFO~38}
\shortauthors{R. Choudhury et al.}

\begin{document}


\title{Triggered  star  formation  and  Young  Stellar  Population  in
  Bright-Rimmed Cloud SFO 38}

\author{Rumpa Choudhury}
\affil{Indian Institute of Astrophysics,
II Block, Koramangala, Bangalore 560 034, India}
\email{rumpa@iiap.res.in}

\author{Bhaswati Mookerjea}
\affil{Department of Astronomy \& Astrophysics, Tata Institute of
Fundamental Research, Homi Bhabha Road, Mumbai 400 005, India}
\email{bhaswati@tifr.res.in}

\and

\author{H. C. Bhatt}
\affil{Indian Institute of Astrophysics,
II Block, Koramangala, Bangalore 560 034, India}
\email{hcbhatt@iiap.res.in}

\begin{abstract}

We  have  investigated the  young  stellar  population  in and  around
SFO~38, one  of the massive globules  located in the  northern part of
the galactic  \HII\ region IC~1396,  using the {\em Spitzer}  IRAC and
MIPS observations  (3.6 to  24 ~\micron) and  followed up  with ground
based  optical photometric and  spectroscopic observations.   Based on
the IRAC and MIPS colors and \ha\ emission we identify $\sim$~45 Young
Stellar  Objects (Classes 0/I/II)  and 13  probable Pre  Main Sequence
candidates.   We  derive the  spectral  types  (mostly  K- and  M-type
stars),  effective  temperatures  and  individual  extinction  of  the
relatively bright and optically visible Class II objects.  Most of the
Class II  objects show variable \ha\  emission as well  as optical and
near-infrared  photometric  variability  which  confirm  their  ``{\em
  youth}".  Based on optical photometry and theoretical isochrones, we
estimate  the spread in  stellar ages  to be  between 1--8~Myr  with a
median age of 3 Myr and a mass distribution of 0.3--2.2~$\msun$ with a
median value around 0.5~$\msun$.  Using the width of the \ha\ emission
line measured  at 10\%  peak intensity, we  derive the  mass accretion
rates   of   individual   objects   to  be   between   $10^{-10}$   to
$10^{-8}$~$\msun$/yr.  From the  continuum-subtracted \ha\ line image,
we  find  that the  \ha\  emission of  the  globule  is not  spatially
symmetric with respect  to the O type ionizing  star HD~206267 and the
interstellar  extinction towards  the globule  is also  anomalous.  We
clearly  detect  an  enhanced  concentration  of YSOs  closer  to  the
southern rim of  SFO~38 and identify an evolutionary  sequence of YSOs
from the rim to the dense core of the cloud, with most of the Class II
objects located at the bright rim. The YSOs appear to be aligned along
two different directions towards the O6.5V type star HD~206267 and the
B0V type star HD~206773.  This is consistent with the Radiation Driven
Implosion  (RDI)  model for  triggered  star  formation.  Further  the
apparent  speed of sequential  star formation  is consistent  with the
speed  of propagation  of shocks  in  dense globules  as derived  from
numerical simulations of RDI.

\end{abstract}


\keywords{stars: low mass, pre-main-sequence, ISM: clouds and \HII\
regions: individual: SFO~38, IC~1396}

\section{Introduction}

Observational and theoretical studies  of star formation over the last
decade  have increasingly  strengthened  the idea  that massive  young
stars play an important role in triggering the formation of subsequent
generations  of  stars.  The  triggers  of  star  formation which  are
typically in the form of wind  and radiation from the massive stars or
expansion  of  \HII\  regions  essentially involve  the  induction  of
compression of  a molecular  cloud externally. Among  several proposed
theories, two models have gathered sufficient observational support in
order to be  regarded as the most plausible  models for triggered star
formation. In the first model  known as the {\em Collect and Collapse}
\citep{elmegreen77,hosokawa06} model, an expanding \HII\ region sweeps
up material into a dense  bordering layer between the \HII\ region and
the molecular  cloud. The compressed  shell of dust and  gas undergoes
fragmentation   and  gravitational   collapse  to   form   new  stars.
Observational  evidence supporting the  ``collect and  collapse" model
includes  detection of  a dense,  fragmented shell  of gas  with newly
formed       stars       surrounding       the      \HII\       region
\citep{deharveng03,deharveng09,zavagno06,zavagno07}.  The {\em collect
  and collapse}  model is however  primarily to explain  the triggered
formation  of massive  stars. An  equally likely  model based  on {\em
  Radiation  Driven   Implosion  (RDI)},  involves   the  creation  of
shock front at the surface  of molecular clouds due to photoevaporation
from the surfaces  of molecular clouds exposed to  the UV radiation of
nearby massive  young stellar clusters.  The  enhanced inward pressure
triggers  the  formation  of  new  protostellar  cores  or  compresses
pre-existing ones  to form a  new generation of  stars.  Observational
evidence for  the RDI model includes typical  spatial distribution and
gradient   in   evolutionary   phases   of   young   stellar   objects
\citep{ikeda08}.   There is  however considerable  ambiguity regarding
whether  RDI actually  induces  the gravitational  collapse or  merely
exposes the young stars by photoevaporation, since these two scenarios
are  observationally  indistinguishable.   Thus  additional  tests  to
ascertain  the temporal  feasibility  of shock  waves  from nearby  OB
associations  actually   triggering  sequential  star   formation  are
required to strengthen the arguments in favor of RDI.

Bright-rimmed clouds  (BRCs) are isolated molecular  clouds located on
the  edges of  evolved \HII\  regions, which  are found  to  have many
signposts of star formation, i.e., {\em IRAS} point sources, molecular
outflows, HH objects  etc. \citet{sugitani99} and \citet{sugitani2000}
observed a sample  of 89 BRCs and found an  excess in the luminosities
and  luminosity-to-cloud ratios  of embedded  {\em IRAS}  sources when
compared with sources in isolated  globules, which is an indication of
enhanced star  formation in the BRCs. Young  stellar objects seemingly
aligned along the axis towards  the ionizing cluster were detected via
near-infrared  (NIR)  imaging of  44  of  these  BRCs. There  is  also
evidence for an age gradient, with older stellar objects closer to the
OB cluster and  the younger objects well inside  the globules, aligned
with the  IRAS sources.  These results are  consistent with sequential
formation of stars while the  shock front advances further and further
into the  molecular cloud.  Thus  BRCs are ideal objects  for studying
and verifying the different models of triggered star formation.

The   \HII\    region   IC~1396,    powered   by   the    O6.5V   type
\citep{stickland95} star HD~206267 in the Trumpler~37 cluster, appears
to sweep up a molecular  ring of radius 12~pc \citep{patel1998} and is
surrounded by 11 Bright-Rimmed clouds with embedded IRAS point sources
\citep{sugitani94,sugitani91}.   SFO~38, also  known  as IC~1396N,  is
located  at  a projected  distance  of  $\sim$11~pc  to the  north  of
HD~206267 with  the bright rim  corresponding to the  ionization front
facing the O6.5V  star.  SFO~38 is located in  the Cep~OB2 association
at  a  distance  of  750~pc  \citep{matthews1979}.  More  recent  {\em
  Hipparcos}  parallax  based measurements  estimate  the distance  to
Cep~OB2 association  to be 615~pc \citep{dezeeuw99}.  However in order
to  make our  results comparable  to most  published results,  in this
paper  we  adopt  a  distance  of  750~pc  for  SFO~38.   The  derived
luminosities  may therefore  be overestimated  by more  than  40\% and
physical   sizes   by   20\%.    The   region   is   associated   with
IRAS~21391+5802,  a  very   young  intermediate-mass  object,  with  a
luminosity of 235~\lsun\  \citep{saraceno96}, which powers an extended
bipolar outflow \citep{sugitani89}.   Based on millimeter observations
\citet{beltran02}  resolved IRAS~21391+5802 into  an intermediate-mass
source  named  BIMA~2  surrounded  by  two less  massive  and  smaller
objects,  BIMA~1  and  BIMA~3.   \cite{valdettaro2005}  detected  {\em
  $H_{2}O$}  maser  emission  at   2.2~GHz  towards  SFO~38  which  is
consistent  with  an  intermediate-mass object.   \citet{neri07}  used
still    higher   angular   resolution    millimeter   interferometric
observations  to reveal  that the  intermediate-mass  protostar BIMA~2
itself  consists  of  multiple  compact  sources.   The  gas  emission
surrounding   IRAS~21391+5802  traces  different   molecular  outflows
\citep{codella01,beltran02,beltran04}.  NIR images  of the region have
also  revealed  a  number   of  small  scale  molecular  hydrogen  and
Herbig-Haro                         (HH)                         flows
\citep{nisini01,sugitani02,reipurth03,caratti06,beltran09}.       These
observations confirm the on going  star formation in the dense core of
SFO~38.

\cite{getman07}  identified  the  mid-infrared (MIR)  ({\em  Spitzer})
counterparts of  X-ray sources detected in  {\em Chandra} observations
and found that  the young stellar objects (YSOs)  are oriented toward
the ionizing star  HD~206267 and show an age  gradient consistent with
the RDI model for triggered star formation.  Recently \cite{beltran09}
studied  the  YSO  population  of  SFO~38  by  obtaining  deep  J,  H,
K$^\prime$   broadband  images   and   deep  high-angular   resolution
observations  in the H$_2$  narrowband filter.   \citet{beltran09} did
not detect any clear NIR excess close to the rim, which is in contrast
to   the  age   1~Myr   of  the   stellar   population  concluded   by
\citet{getman07}.  \citet{beltran09}  suggest that the  YSOs closer to
HD~206267 could  actually be  younger than the  Class II  objects they
appear  to be  since  their circumstellar  environment were  disrupted
completely   by    intense   UV   radiation   field    from   the   OB
association. \citeauthor{beltran09}  thus suggest that  in general the
apparent  age  sequence  seen   close  to  OB  associations  need  not
necessarily be  the actual  evolutionary sequence and  in the  case of
SFO\,38 there is no concrete  evidence of star formation by the nearby
OB association.

In this paper  we address the controversial issue  of the distribution
of YSOs  by making use of  the extremely sensitive  {\em Spitzer} IRAC
(3.6 to 8.0~\micron) \& MIPS  (24~\micron) data in order to probe {\em
  all} embedded YSOs and protostars  in SFO~38.  This thus extends the
work     by    \citet{getman07},     who     had    considered     the
3.6--5.8~\micron\ characteristics  only of those  MIR sources
from  which  X-ray emission  had  been  detected.   Our approach  also
complements the  work by \cite{beltran09} which  makes use of
only the  NIR observations to decipher  any age gradient  of the YSOs.
In  this  work we  shall  show  that most  of  the  Class 0/I  sources
identified  in  the  MIR   are  either  barely  detected  (with  large
uncertainties in fluxes) or not detected  at all in the NIR and in the
X-rays.  In this paper  we have completed the multiwavelength overview
of SFO~38  by obtaining optical  broadband images in {\em  BVRI} bands
and narrow band \ha\ image to understand the detailed structure of the
bright-rim. Further in order to confirm the ``{\em youth}" of the YSOs
identified  in  the  MIR  we   have  made  use  of  medium  resolution
spectroscopy  of the  optically visible  YSOs in  \ha\  line emission.
Observation details of  { \em Spitzer} and optical  {\em BVRI} imaging
and medium resolution  spectroscopy are presented in \S~\ref{sec_obs}.
We describe the  selection of YSO candidates based  on MIR color-color
diagram  (CCD)  and the  detailed  analysis  of  optical--MIR data  in
section \S~\ref{sec_results}.  In \S~\ref{sec_dis} we discuss
the probable  star formation scenarios  and in \S~\ref{sec_conclusion}
we summarize the results.

\section{Observations and Data Reduction
\label{sec_obs}}
\subsection{{\em Spitzer} IRAC \& MIPS  observations}

We have extracted  IRAC (3.6, 4.5, 5.8, \& 8~\micron)  and MIPS (24 \&
70~\micron)  observations from the  Spitzer Space  Observatory archive
(Program ID 30050: Star Formation  in Bright Rimmed Clouds by Fazio et
al.).  The IRAC  data were taken in the High  Dynamic Range (HDR) mode
using  a   single  AOR  (Astronomical  Observation   Request)  with  a
five-point dither pattern.  We have processed both the short (0.6~sec)
and the  long (12~sec) integration Basic Calibrated  Data (BCD) frames
in each  channel using the  Artifact mitigation software  developed by
Sean                      Carey                     \footnote{
    \url{http://web.ipac.caltech.edu/staff/carey/irac_artifacts/index.html}}
and  created  mosaics  using  MOPEX.   In all  IRAC  bands  we  detect
point-sources down to 70~$\mu$Jy.  We have also created mosaics of the
MIPS  24 and  70~\micron\ BCDs  using MOPEX.   Both data  sets  are of
reasonably good quality. The 70 $\mu$m-image, due to the lower angular
resolution and  smaller area mapped, shows only  one bright point-like
source embedded in the globule.

We  have carried  out multiframe  PSF photometry  using the  tool APEX
developed  by Spitzer  Science  Center(SSC) on  all the  {\em
  Spitzer}  IRAC  images  and  on  the MIPS  images.     The  long
  integration  3.6   and  4.5~\micron\  IRAC  images   show  signs  of
  saturation  on  bright  stars.    For  the  saturated  sources,  the
  photometry derived from the  short integration images were used.  It
  is  difficult  to  disentangle  the  sources in  regions  of  strong
  emission from  the associated  Photon Dominated Regions  (PDRs) from
  the surrounding clouds and  to derive their accurate photometry.  We
  have used a combination  of automated routines and eye-inspection to
  detect sources and extract photometry of these sources from the IRAC
  and  MIPS  images.   For   sources,  which  APEX  failed  to  detect
  automatically at one  or several wavelengths, we have  used the user
  list  option in APEX  to supply  the coordinates  for the  source to
  successfully derive a PSF fit.  This enabled us to derive photometry
  for  every source  which we  could visually  identify on  any image.
  Following  the IRAC  and MIPS  Data  Handbooks we  have adopted  the
  zero-points for conversion between  flux densities and magnitudes to
  be 280.9, 179.7, 115.0, 64.1 and 7.14 Jys in the 3.6, 4.5, 5.8, 8.0,
  and 24\,\micron\ bands, respectively.

\subsection{Optical {\em BVRI} Photometry}

SFO~38 was observed on 17th  July, 23rd August and 15th September 2009
using   Bessell's   broad  band   filters   {  \em   VR}~(300\mult8~s,
300\mult8~s),  {\em BVI}~(900\mult8~s,  300\mult8~s,  150\mult8~s) and
{\em BI}~(900\mult8~s, 120\mult12~s)  respectively and on 4th November
2008 using the narrow band {\em \ha} ~(900~s) with the Himalayan Faint
Object Spectrograph Camera (HFOSC) mounted on the 2m Himalayan Chandra
Telescope  (HCT)  of   the  Indian  Astronomical  Observatory,  Hanle,
India\footnote{ \url{http://www.iiap.res.in/centers/iao}}.  HFOSC uses
the central 2K\mult2K region of  the 2K\mult4K CCD in imaging mode and
covers  a  field  of  $10^{'}$\mult$10^{'}$  with  a  plate  scale  of
0\farcs296~ $pixel^{-1}$.  The nights were photometric with an average
seeing of  1\farcs5 to 1\farcs8.   \cite{landolt} photometric standard
stars were observed  on all the nights to  calibrate the target stars.
Data  were  reduced  using   standard  tasks  available  within  Image
Reduction and  Analysis Facility (IRAF)\footnote{The  IRAF software is
  distributed  by  the National  Optical  Astronomy Observatory  under
  contact        with        National       Science        Foundation.
  \url{http://iraf.noao.edu/}}.  Bias subtracted, flat field corrected
and aligned  frames were  combined to make  the master frame  for each
filter.   Astrometric calibration  was  applied to  the master  frames
using  the  IDL  procedure  {\em  STARASTT} of  IDL  Astronomy  User's
Library\footnote{  \url{http://idlastro.gsfc.nasa.gov/}}.  Photometric
magnitudes  were calculated  by aperture  photometry with  the optimal
aperture  adopted as  the radius  where the  difference  in magnitudes
between two  consecutive apertures is  less than 1$\%$.   The limiting
magnitude  is defined  as the  magnitude at  which the  mean magnitude
error of the star becomes 0.1 mag which implies a 10$\sigma$ detection
corresponds  to a  signal-to-noise ratio  (S/N) of  10.   The limiting
magnitudes were V:~21,  R:~21 on 17th July ,  B:~22, V:~22, I:~19.5 on
23rd  August  and  B:~22.2,   I:~20.2  on  15th  September.   Aperture
corrections for each  frame were derived from the  bright and isolated
stars  and applied  to the  faint  stars.  The  standard deviation  of
residuals  of observed and  transformed magnitudes  and colors  of the
standard stars  are within  the range of  0.01-0.02 mag.  We  used the
broadband  {\em R}  filter image  for continuum  subtraction  from the
narrow  band   \ha\  image  of   SFO~38.   Based  on  the   recipe  of
\citet{waller1990}, the  point spread functions of the  R-band and the
narrow  band \ha\ images were matched and the images were scaled before
subtraction to get the \ha\ emission line image of SFO~38.

\subsection{Medium Resolution Spectroscopy}

Medium resolution ($\sim$7~\ang)  spectra were obtained for relatively
bright and  optically visible YSOs during July--November  of both 2008
and 2009 in the  wavelength range (5200--9200~\ang) with the Himalayan
Faint  Object Spectrograph  Camera  (HFOSC) mounted  on the  Himalayan
Chandra Telescope  (HCT).  Typical exposure time of  each spectrum was
3600~s.   The  spectra were  bias  subtracted,  flat  fielded and  the
one-dimensional  spectra were  extracted using  the standard  tasks of
Image Reduction and Analysis Facility  (IRAF). The arc lamp spectra of
FeNe were used for wavelength calibration.  We further used the strong
night sky emission lines  e.g.  [OI] $\lambda$5577 \ang, $\lambda$6300
\ang\  lines and rotational  and vibrational  bands of  OH in  the red
region  of the individual  target spectrum  to improve  the wavelength
calibration  and achieved  an  accuracy of  $\pm$~0.5  \ang\ for  each
target star.

\section{Results and Analysis
\label{sec_results}}

\subsection{Mid-Infrared (3.6 to 24~\micron) view of SFO~38}

\begin{figure*}[ht]
\begin{center}
\resizebox{\hsize}{!}{\includegraphics[angle=0,width=18.0cm,angle=0]{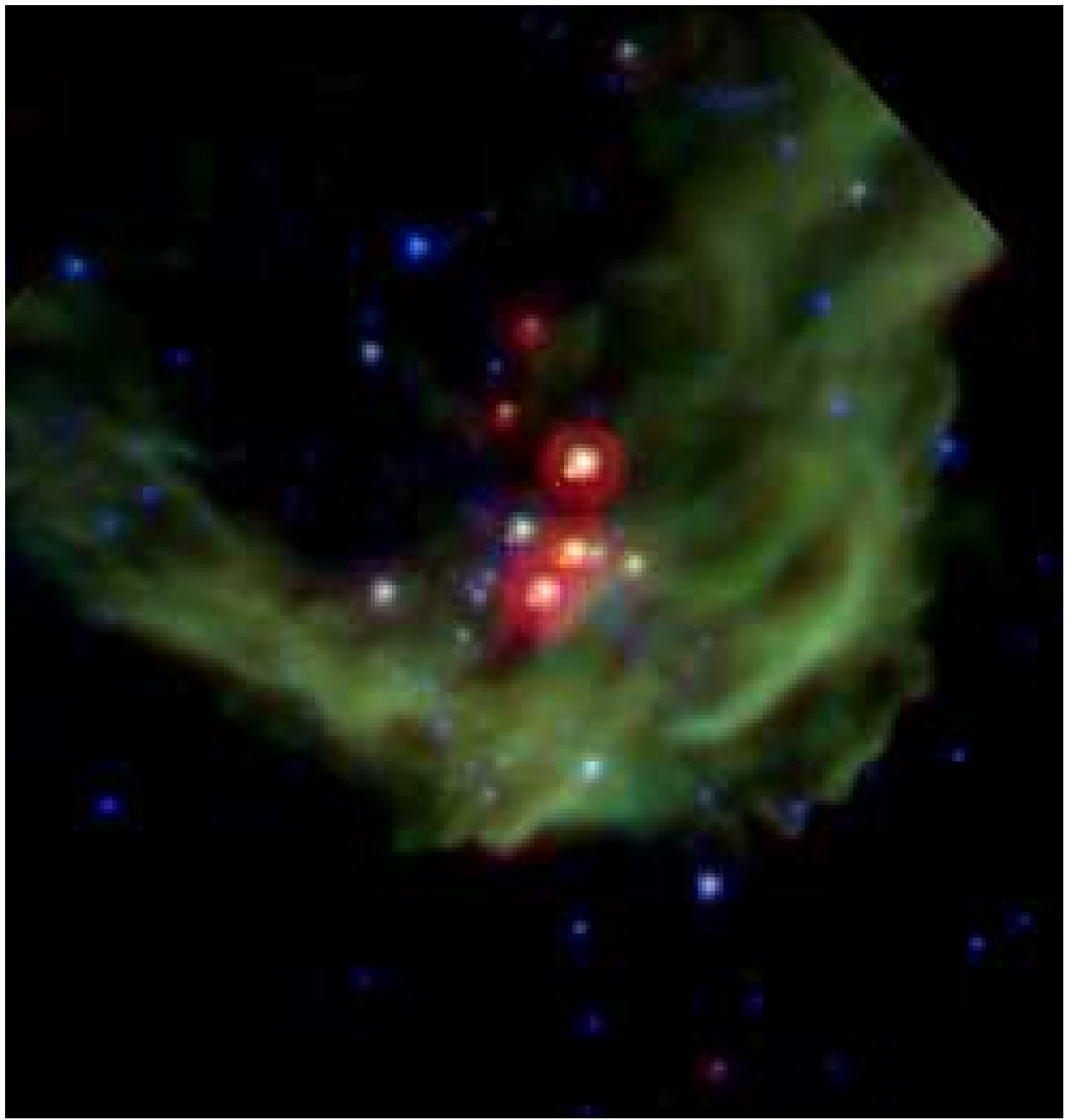}}
\caption{IRAC-MIPS color-composite image of SFO 38 using 3.6~\micron\ (Blue), 8~\micron\ (Green) 
and 24~\micron\ (Red). The image is centered at $\alpha_{2000}$  = 21$^h$40$^m$42$^s$ and
$\delta_{2000}$ = 58\arcdeg16\arcmin10\arcsec and extends over 5\arcmin$\times$5\arcmin\ ($\alpha\times\delta$).}
\label{fig_iracyso}
\end{center}
\end{figure*}

Figure~\ref{fig_iracyso}  presents the  three-color  image of  SFO~38
using the  IRAC 3.6,  8.0 and  the MIPS 24~\micron\ bands.  The image
shows  that  the  emission from  the  front  side  of the  globule  is
dominated by strong  polyaromatic hydrocarbon emission (PAH-emission),
and  rather clumpy.  This  could  in part  be  due to  Rayleigh-Taylor
instabilities  at the  ionization front,  but it  is also  due  to the
outflows from  the young stars. The  cometary shape of  the globule is
quite apparent with  the tip pointed towards the O  star HD 206267 and
the eastern edge  appears to be more abruptly  truncated to the south.
This indicates  that the  south-eastern edge of  the globule  has most
likely experienced intense ionizing  radiation field, which has eroded
the cloud material more than  on the western side. Comparison with the
\ha\ emission line image (Fig.\,\ref{fig_ysomap}) also confirms the
presence of a more pronounced ionization front to the south-east and a
rather tenuous layer of ionized gas to the west of SFO~38.

The aim  of this paper is  to identify the young  and embedded stellar
population in  SFO~38.  While  the 3.6 and  4.5~\micron\ IRAC  data is
particularly sensitive, a large  number of stellar objects in addition
to  the YSOs  also detected  in  these bands.  We have  thus used  all
sources  detected  either  at  5.8  or  at  8.0~\micron\  or  at  both
wavelengths to create a list  of MIR sources in SFO~38. Photometry for
all these  sources, if  detected, are then  extracted from the  3.6 \&
4.5~\micron\  images.  We  thus detect  98,  106, 106  and 98  sources
respectively in the IRAC 3.6,  4.5, 5.8 and 8.0~\micron\ images and 14
and 1 sources respectively in  the MIPS 24 \& 70~\micron\ images.  All
the sources detected at 24~\micron\ are found to have been detected in
the IRAC  bands. In all  we identify a  total of 110 MIR  sources.  We
have  crosscorrelated the  MIR sources  with  the NIR  sources in  our
mapped region  from the 2MASS point  source catalogue, as  well as the
sources detected in the NIR by \citet{nisini01} and \citet{beltran09}.
We have used the following  association radii : 1\arcsec\ for the IRAC
images, 2\farcs5 for the MIPS  24~\micron\ image and 2\arcsec\ for NIR
data.  Table~\ref{tab_mirsrc} gives the coordinates of the 110 sources
identified in SFO~38 together with their NIR magnitudes, {\em Spitzer}
IRAC and MIPS flux densities and a preliminary classification based on
selected   color-color   plots  and   other   criteria  described   in
Sec.~\ref{sec_ysoclass}.  In  Table~\ref{tab_mirsrc} we have  used the
prefix SFO38,  but throughout  the paper we  refer to them  as MIR-nn,
where nn is the number of the  source.  Out of the 110 sources 80 were
found  to have NIR  counterparts.  However  10 sources  out of  the 80
sources with  NIR counterparts were  found to coincide with  the H$_2$
knots identified by \citet{beltran09}.

Additionally in  order to identify  the stellar and pre  main sequence
(PMS)  stars  we used  photometry  extracted  from  the IRAC  3.6  and
4.5~\micron\ long integration images.  The 3.6 and 4.5~\micron\ images
are more sensitive, have a cleaner PSF, and appear to be less affected
by  nebular emission  than the  8~\micron\ image.   We  identified 161
additional  sources,   which  are  detected   in  both  the   3.6  and
4.5~\micron\ wavebands.   Of these sources  113 are found to  have NIR
counterparts, and 6 sources are coincident with H$_2$ emission knots.

\subsection{Classification of  YSOs based on Near- and Mid-infrared colors
\label{sec_ysoclass}}

In  regions  of high  obscuration  like  SFO~38,  which has  a  visual
extinction  of up  to 20~mag,  mid-infrared color  indices  provide an
ideal tool for  identifying YSOs and classify them  according to their
phases of  PMS evolution \citep{getman07}.  Here we  use several color
indices  generated using  the  {\em  Spitzer} IRAC  and  MIPS data  to
characterize the  evolutionary stages of  the MIR sources  detected in
SFO~38.

\citet{stern05} demonstrated that (a) normal star-forming galaxies and
narrow-line  AGNs  with  increasing   5.8  and  8.0~\micron\  and  (b)
broad-line AGNs with red, nonstellar  SEDs, result in colors which are
very similar to bona-fide YSOs.  Thus prior to identifying PMS sources
based on the color indices it is necessary to inspect deep IRAC images
for    contamination    due     to    the    extragalactic    sources.
\citet{gutermuth09}  have  extensively   discussed  the  criteria  for
identifying  such  extragalactic   objects  in  the  IRAC  color-color
diagrams. Using the criteria given  in the appendix of their paper, we
identify MIR-2, 17,  68, 70, 75, 81,  84, 89, 101 and 103  as AGNs and
MIR-77, 81, 93 and 103 as galaxies strongly emitting in the PAH bands.
Of  these sources,  MIR-17, 89  and 103  satisfy most  of the
criteria for  extragalactic objects.  The  profiles of sources  in the
$R$-band is typically quite a  good indicator of whether the source is
actually extragalactic or stellar.   However all the sources in SFO~38
which satisfy the criteria mentioned above are extremely faint so that
with  the exception  of MIR-2,  these sources  though visible  are not
detected clearly enough in the  $R$-band so that the intensity profile
can be studied.  Further, none of  the sources is detected in the MIPS
24~\micron\ image either.  The $R$-band profile of MIR-2 appears to be
stellar with  a faint companion on  one side and  MIR-84 is associated
with an \ha\  emission star, so we do not  identify both these sources
as  extragalactic.   In the  absence  of  any  conclusive evidence  we
presently assume  the sources identified by the  empirical criteria as
above to be extragalactic.

The 4.5~\micron\ IRAC  band of {\em Spitzer} covers  many of the H$_2$
pure rotational lines, as a result  of which this band records a large
excess of emission at the  position of the unresolved blobs of shocked
emission from high velocity outflows which interact with the molecular
cloud.  Using  the empirical criteria  outlined by \citet{gutermuth09}
here  we flag  the  sources MIR-24,  27,  53, 65  and  98 as  possible
features due  to shocked H$_2$ emission. Based  on 2.12~\micron\ H$_2$
integrated line  emission \citet{beltran09}  identified as many  as 97
emission  features  possibly created  by  H$_2$  excitation by  shocks
driven by outflows powered by YSOs. We find that 10 out of the 110 MIR
sources that we have identified  are associated with such H$_2$ knots.
These include the  sources MIR-20, 24, 27, 35, 53, 65,  74, 75, 98 and
105. Thus the color indices for 5 out of the 10 sources are consistent
with  the observed association  with H$_2$  knots.  The  source MIR-75
which also satisfied the criteria for being an extragalactic object is
thus identified as a H$_2$ knot.

Figure~\ref{fig_iracmipscol}  presents  color-color  diagrams of  NIR,
IRAC and MIPS  24~\micron\ sources detected in the  SFO~38 globule. We
have  used  several criteria  (shown  as  dashed  lines and  boxes  in
Figure~\ref{fig_iracmipscol})  to identify  the  potential YSOs  using
these color-color diagrams.

The most stringent classification  scheme uses the IRAC ([3.6]--[5.8])
colors  and the  [8]--[24] IRAC  and MIPS  color. At  24  \micron\ the
reddening due to  extinction is small and the  photospheric colors are
very  close  to  zero  for  all  spectral  types  \citep{muzerolle04}.
Therefore the  [8]--[24] color is  very sensitive to  infrared excess,
but of course not all young  stars are bright enough to be detected at
24 \micron.   Using this color-color diagram we  find fourteen sources
with infrared excess  (Figure~\ref{fig_iracmipscol} ({\em left})).  We
identify MIR-34, 36, 48, 50, 54, 55 and 59 as Class 0/I sources, while
MIR-11,  31, 38, 45,  82 and  86 are  within the  Class II  regime and
MIR-29 which has a large [8.0]--[24] color excess and very small value
for the [3.6]-[5.8]  color is identified as a  transition disk object.
\citet{joergensen07}  predicted  sources  with [3.6]--[4.5]$>$1.0  and
      [8]--[24] $>$ 4.5 to be YSOs driving molecular outflows. We find
      that the sources  MIR-34, 48, 50, 54, and 59,  all of which have
      been identified  as Class 0/I objects also  satisfy the criteria
      for being sources driving outflows.

Figure~\ref{fig_iracmipscol}~({\em middle})  presents the [3.6]--[4.5]
vs [5.8]--[8.0] color-color  plot for the sources detected  in all the
IRAC  bands.  Sources  with  the colors  of  stellar photospheres  are
centered  at ([3.6]--[4.5],[5.8]--[8.0])=(0,0) and  include foreground
and background stars as well as diskless (Class III) pre-main sequence
stars.   Sources which  satisfy the  criteria  $-0.1\leq$ [3.6]--[4.5]
$\leq0.1$ and  $0.2\le$ [5.8]--[8.0]  $\leq1.1$ are classified  as the
transitional disk  objects \citep{fang09} which  also do not  have any
NIR  color excess.  The  box outlined  in Figure~\ref{fig_iracmipscol}
({\em   middle}),   defines   the   location  of   Class   II   objects
\citep{megeath04,allen04}, i.e. sources  whose colors can be explained
by young, low-mass stars  surrounded by disks.  \citet{hartman05} have
shown  from their  observations of  young stars  in  the Taurus-Auriga
complex  that  Class  0/I  protostars require  [3.6]--[4.5]$>0.7$  and
[5.8]--[8.0]$>0.7$.  With the exception of MIR-29, the classes derived
for the sources identified in the MIPS 24~\micron\ band and plotted in
Fig.~\ref{fig_iracmipscol}({\em left})  are completely consistent with
the classes derived from  the [3.6]--[4.5] vs [5.8]--[8.0] color-color
plot. We identify 10 of the MIR sources as being associated with H$_2$
knots, this  includes MIR-75,  which otherwise satisfies  the criteria
for  extragalactic objects.  We also  plot NIR  and {\em  Spitzer} MIR
color-color  diagram  of the  identified  YSOs  and transitional  Disk
objects in Figure~\ref{fig_iracmipscol}~({\em right}).

Table~\ref{tab_mirsrc} presents a summary of classification of the MIR
sources  based on the  color-color diagrams  presented here.   We have
additionally looked into  the [3.6]--[5.8] vs [4.5]--[8.0] color-color
diagram  and found  that the  Classes  derived from  this diagram  are
consistent with the classes  derived based on the color-color diagrams
presented here. For most sources it was possible to arrive at a unique
Class based  on all  three color-color diagrams.  For all  sources the
Class derived based on at least two color-color diagrams match.

Use of the more sensitive long integration 3.6 and 4.5~\micron\ images
in  combination with  2MASS  data (following  a classification  scheme
involving the  $K_{\rm s}$--[3.6]  and [3.6]--[4.5] colors)  yields an
additional 113 sources,  11 of which are PMS  objects.  Since the long
integration  3.6  and  4.5~\micron\  images  go deeper  and  are  less
affected by extinction than the 2MASS survey, we also checked how many
of these IRAC  sources which have no 2MASS  associations are likely to
be  PMS stars.   Among  the 48  remaining  sources which  have no  NIR
counterparts, following  the arguments of  \citet{mookerjea2009} if we
use the color criterion, [3.6]-[4.5]  $\geq$ 0.2, we obtain 8 sources.
However  6 of  these  8  sources are  actually  associated with  H$_2$
emission      knots     as     detected      by     \citet{beltran09}.
Table~\ref{tab_nirpms} lists  the 13 PMS candidates  identified on the
basis of NIR and 3.6 and 4.5~\micron\ magnitudes.

Excluding the  H$_2$ knots and  the possible extragalactic  objects we
identify  10 Class  0/I  sources, 3  Class  I sources,  13 Class  I/II
sources (occupying the top-left and bottom-right corners), 14
Class II sources  and 19 transitional disk objects  in SFO~38. Thus we
find  a  total  of 40    YSOs  (Classes  0/I/II) and  13  PMS
candidates  based on MIR  color indices.  We note  that 13  Class I/II
objects  and 19  Transitional Disk  objects are  $\sim$~30$\%$  of the
selected   YSOs  and   show  noticeable   color  excess   only  beyond
6~\micron.

\subsection{\ha\ emission of SFO 38 and spatial distribution of YSOs}

\begin{figure*}
\begin{center}
\resizebox{\hsize}{!}{\includegraphics{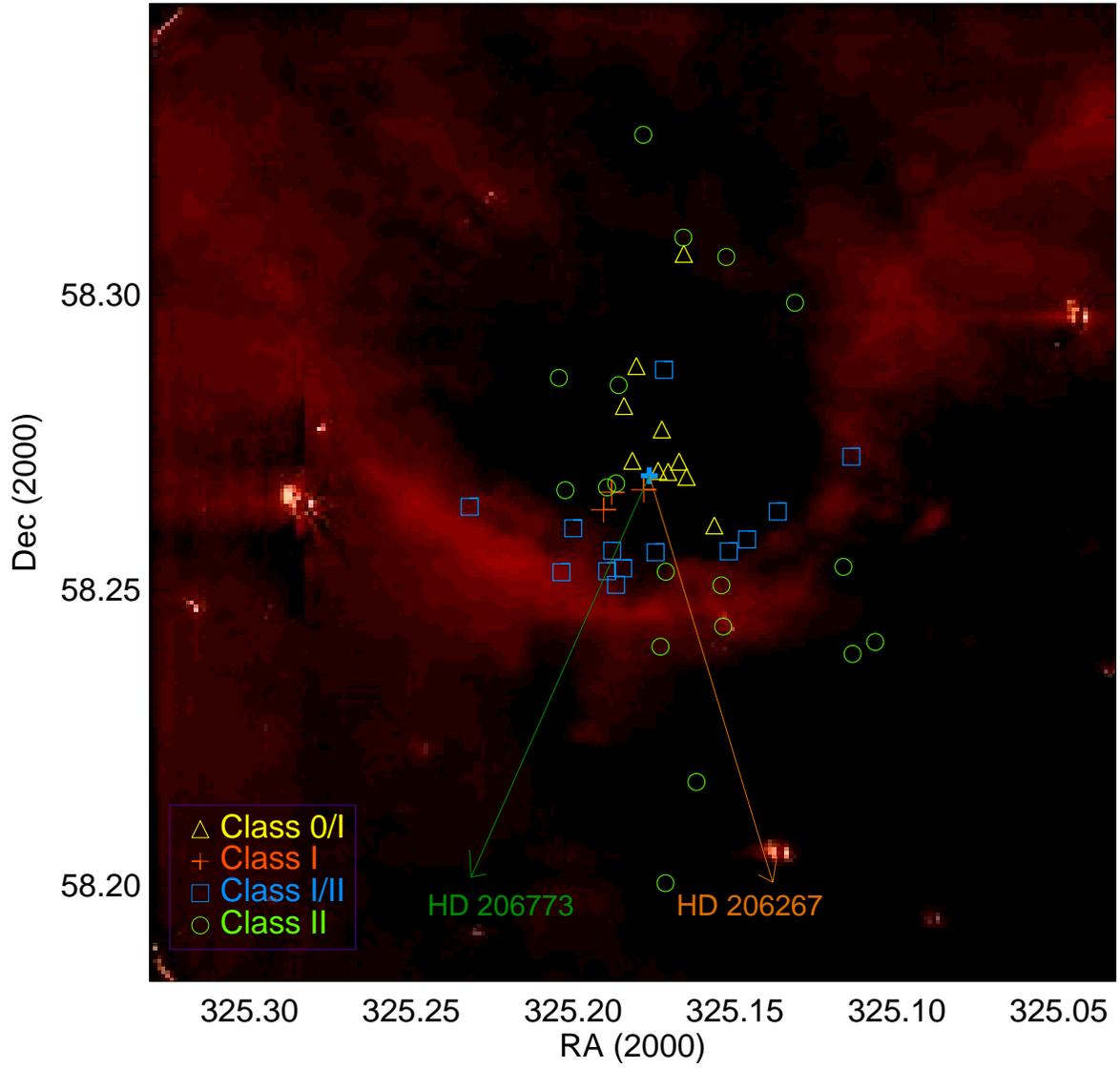}}
\caption{Continuum subtracted \ha\ emission line image of SFO 38. Class 0/I to Class II
 YSOs are overplotted with different symbols. The directions towards HD 206267 and HD 206773 are marked with two different {\em arrows}.
\label{fig_ysomap}}
\end{center}
\end{figure*}

\begin{figure*}
\begin{minipage}{5.5cm}
   \includegraphics[width=5.4cm]{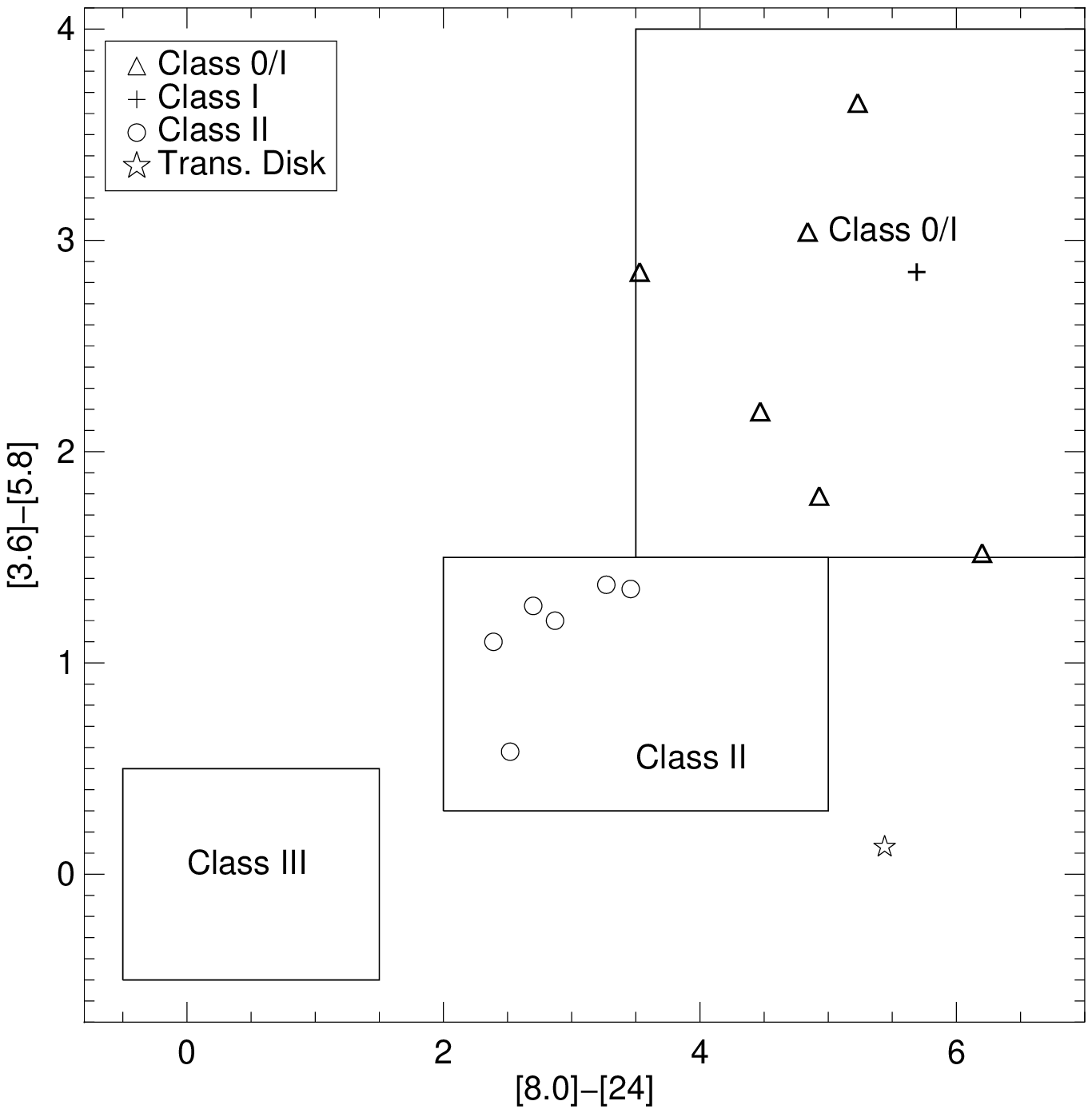}
  \end{minipage}
\begin{minipage}{5.5cm}
   \includegraphics[width=5.4cm]{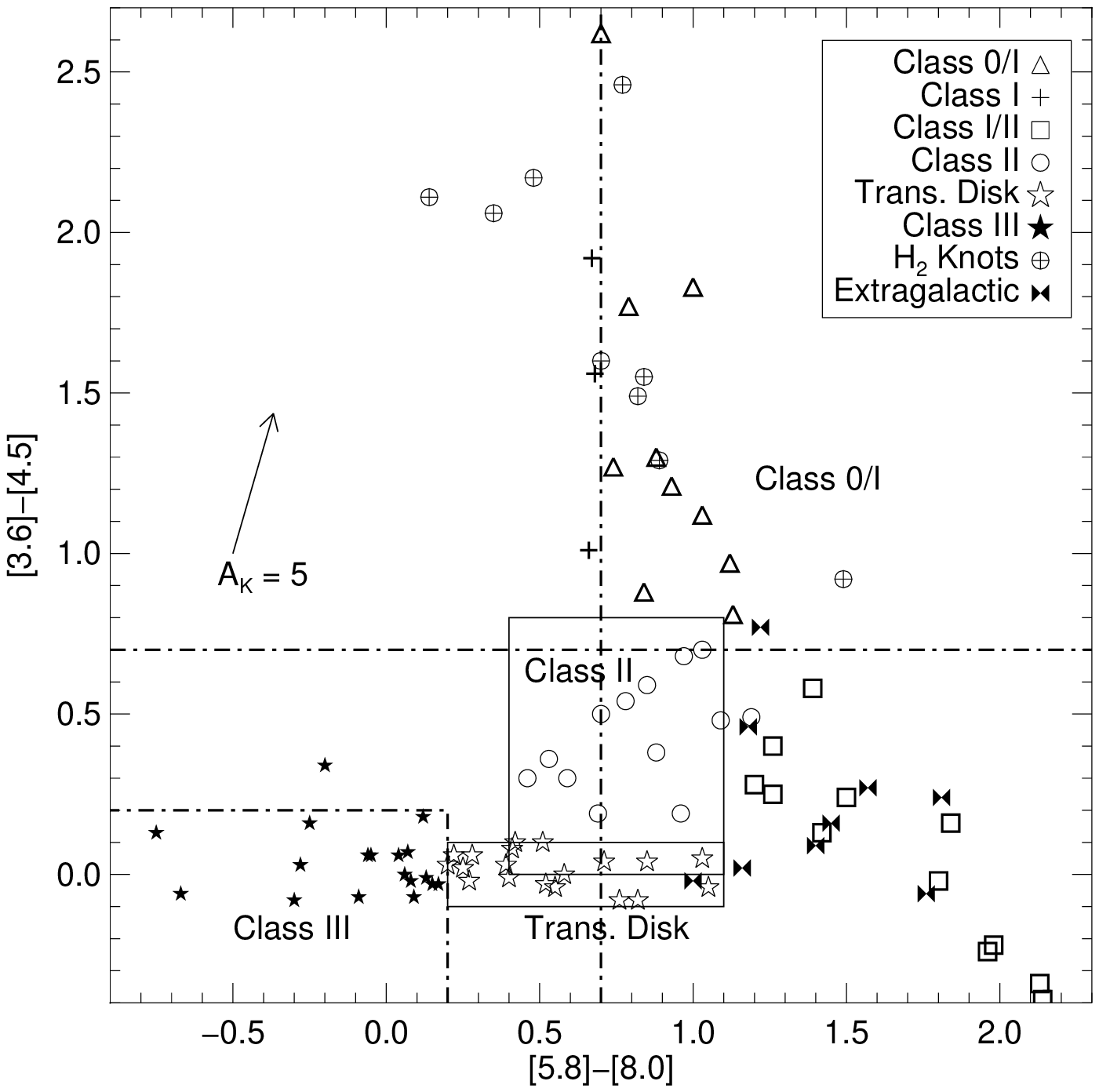}
  \end{minipage}
\begin{minipage}{5.5cm}
   \includegraphics[width=5.4cm]{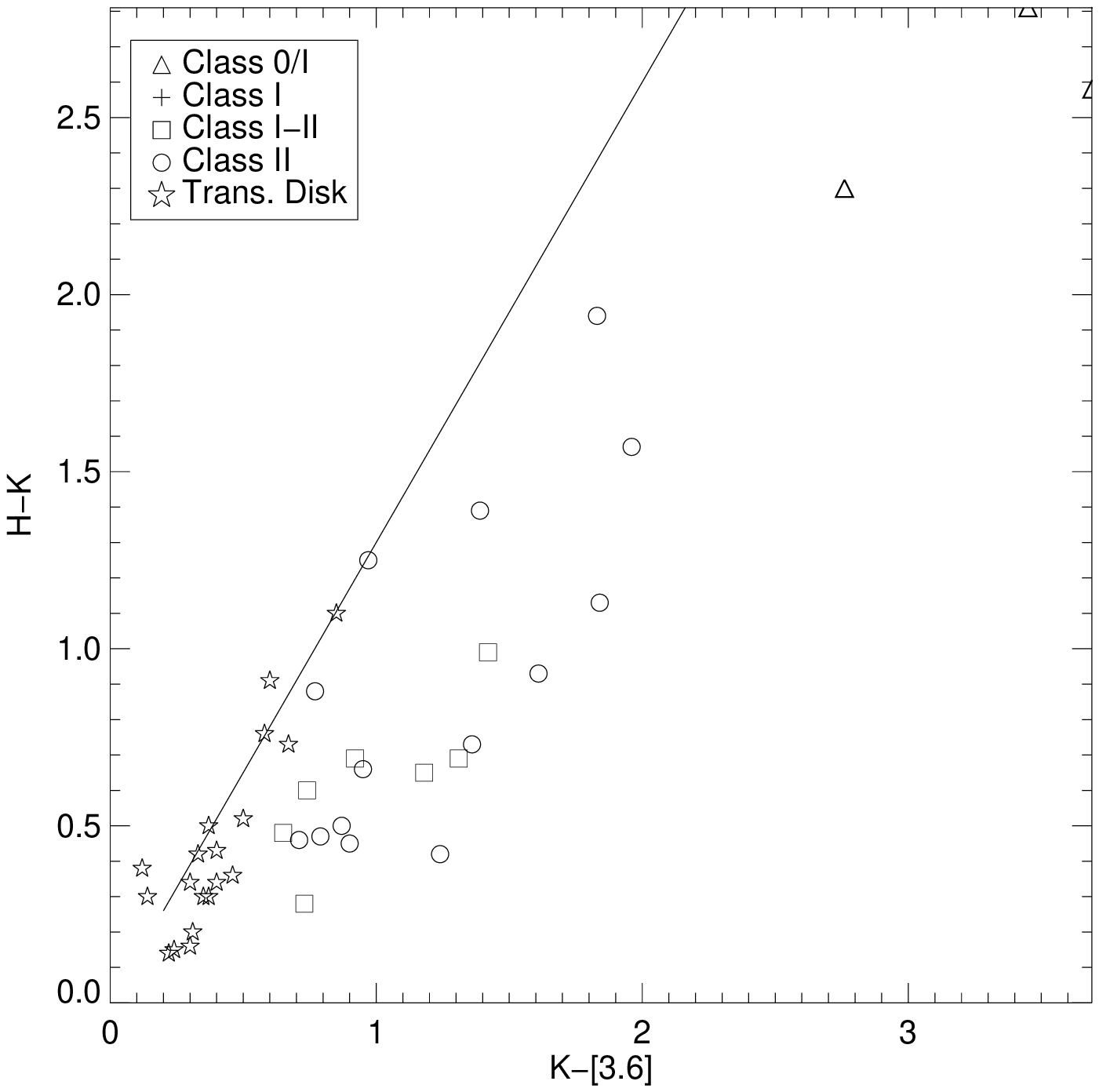}
  \end{minipage}
\caption{Color-color  diagrams  for   all  MIR   sources.
  Approximate classification  criteria adopted from \citet{megeath04},
  \citet{fang09} and \citet{muzerolle04} are  shown in the two panels.
  (a) MIR color-color  diagram based on {\em Spitzer}  3-band IRAC and
  MIPS  photometry and  (b) 4  band IRAC  color-color  diagram. Dashed
  lines  in the two panels are  taken from  \citet{hartman05}.  Reddening
  vector  corresponding to  the extinction  laws given  by  the fitted
  function from \citet{indebetouw05} is shown in {\em (b)}. The dashed
  lines [3.6]--[4.5] = 0.7, [4.5]--[5.8]  = 0.7 and [5.8]--[8.0] = 0.7
  discriminate  Class   II  sources   from  Class  I/0   sources,  and
  [3.6]--[4.5]  =  0.2, [4.5]--[5.8]  =  0.2  and  [5.8]--[8.0] =  0.2
  discriminate Class III from Class II and Transitional Disk sources. (c) NIR and {\em Spitzer} MIR color-color diagram. {\em Solid} line is the reddening vector due to interstellar extinction \citep{tapia81}. Sources to the right of the reddening vector are YSOs with NIR excess.
\label{fig_iracmipscol}}
\end{figure*}

\cite{sugitani91} classified SFO~38  as a {\em B} type  BRC powered by
O6.5 star HD~206267.   The bright rim structure is  quite prominent in
the   continuum   subtracted   \ha\    line   image   of   SFO~38   in
Figure~\ref{fig_ysomap}.    The  \ha\   emission  nebulosity   of  the
bright-rim of the globule is  asymmetric with respect to the geometric
axis of the  globule as well as the direction  of the adopted ionizing
source HD~206267. The  ionized rim appears to be  brighter towards the
east and  rather diffuse towards  the western edge,  facing HD~206267.
The \ha\ intensity distribution of the diffuse material in between the
bright-rim and the ionizing star  also show similar kind of asymmetric
distribution as  that of the  bright-rim. We, therefore,  searched for
other massive  O and B type  stars in the surrounding  region that can
produce the relatively brighter \ha\  emission on the left side of the
globule.  We find a B0V type \citep{jaschek82} star HD~206773 which is
situated on the symmetry axis of  the left side of the bright-rim at a
projected  distance of  $\sim$ 7.6~pc  from  SFO~38.  \cite{dezeeuw99}
tagged  HD~206773 as  a member  of Cep~OB2  association.  The
distance  towards  the  star  is between  550--1000~pc,  adopting  the
parallax measurement  and the associated  error \citep{hipnew07} which
is consistent  with the  distance of SFO~38.   The number  of ionizing
photons  emitted per  second  by  a B0V  star  is $\sim$  10$^{48.02}$
\citep{photonrates03} and using n$_{H} \simeq 10^{7}$~$m^{-3}$
  for number density  of hydrogen in the interior  of the \HII\ region
  IC~1396,  the Str\"{o}mgren  radius  would be  $\sim$ 8.5~pc  which
suggests  that HD~206773  could  be a  potential  ionizing source  for
SFO~38.  Fig.~\ref{fig_ysomap} shows the  distribution of the YSOs and
the directions toward HD~206267 and  HD~206773 on the \ha\ line image.
We find  that most of  \ha\ emitting Class  II objects lie  within the
bright-rim.   Class I/II  objects  are situated  behind  the Class  II
objects  in the  intermediate region  between the  bright-rim  and the
dense portion of  the globule.  All the Class 0/I  and Class I objects
are situated well within the dense  core part of the globule. There is
no Class  0/I to Class I/II  YSO which is situated  before the ionized
rim.  The YSOs appear to  have elongated distribution to the right and
left sides of the bright rim with two different axes of elongation.

\subsection {Additional YSO candidates based on \ha\ emission 
\label{sec_ysohalpha}}

\ha\  emission surveys  are  very  useful to  identify  low mass  ($<$
2~M$\sun$)  Classical  T Tauri  stars  (CTTS)  in  young star  forming
regions as the \ha\ emission  is due to the accretion of circumstellar
material  in  YSOs.  These  observations  also  detect  a fraction  of
Weak-lined T Tauri stars with  weak \ha\ emission which is believed to
arise  from  enhanced   chromospheric  activities.   Based  on
  \ha\  emission,  we  searched  for  the  additional  YSO  candidates
  available  in the  literature  outside  the field  of  view of  {\em
    Spitzer}  observations  to  complement  the list  of  YSOs  around
  SFO~38.  Many of the candidate YSOs in and around SFO 38 were found
during \ha\ objective prism survey by \cite{ogura02,ikeda08}.  A total
of 14 out of the 17 \ha\ emission line sources \citep{ogura02,ikeda08}
are situated within  the field of view of IRAC.   We consider the rest
of     the    \ha\     emission    line     sources     {\em    viz.,}
\object{[OSP2002]~BRC~38~1},     \object{[OSP2002]~BRC~38~16},     and
\object{[OSP2002]~BRC~38~17}  as  YSO  candidates.   We  also  find  4
\ha\  emission   line  sources  near   SFO~38  in  the   catalogue  of
point-source  \ha\  emission-line objects  selected  from the  INT/WFC
Photometric  \ha\  Survey  (IPHAS)  of  the  northern  Galactic  plane
\citep{iphas}.   We   identify  \object{IPHAS~J214027.32+581421.3}  as
\object{[OSP2002]~BRC~38~2},   \object{IPHAS~J214036.90+581437.9}   as
\object{[OSP2002]~BRC~38~6},   \object{IPHAS  J214041.23+581158.5}  as
\object{[GFG2007]~64}.   Thus, we also  consider \object{[GFG2007]~64}
and \object{IPHAS~J214042.81+581937.4} as YSO candidates.

\subsection{Optical and NIR variability}

Out of the  115 sources (110 MIR + 5  \ha\ emission sources) mentioned
in       Secs.~\ref{sec_ysoclass},       ~\ref{sec_ysohalpha}      and
~\ref{sec_medspec} we detect  a total of {\em 39}  sources in {\em B},
{\em 46 } sources in {\em V}, {\em 51} source in {\em R} and {\em 63 }
source in {\em I} bands.  The {\em BVRI} magnitudes of all the sources
detected in {\em V}  band are presented in Table~\ref{tab_optmag}.  We
do not detect the optical counterparts of the 13 PMS candidates listed
in  Table~\ref{tab_nirpms}.  Photometric  variability,  a very  common
characteristic of  YSOs, can happen due to  various physical processes
{\em viz.,}  variation in stellar spots or  variable obscuration which
affect the  emission from  its photosphere.  Since  we have  {\em BVI}
observations separated by several months,  we have also looked for the
variability in  the {\em BVI} bands  for all the  sources.  The median
variability  in {\em  BVI} bands  are $\sim$  0.04 mag  which  is also
indicative of  the accuracy  of the optical  photometric measurements.
We consider a source to  be {\em variable} if the magnitude difference
is  $\geq$  0.1  mag  (Table~\ref{tab_optmag}) at  the  two  observing
epochs.  MIR-5,  MIR-32, MIR-43, MIR-45  and MIR-49 show  variation in
{\em   BVI}  bands.    In  particular   MIR-5   with  W$_\lambda$(\ha)
$\sim$~80~\AA\  shows  $\sim$~1 mag  variability  in  {\em  B} over  a
timescale of month.

YSOs also  show variation in NIR bands  which may or may  not have any
correlation  with  the optical  variation  \citep{eiroa02}.  Both  the
stellar photosphere  and the circumstellar disk contribute  to the NIR
fluxes of PMS stars.  Hence, NIR  variability can be either due to the
same  physical processes  responsible for  optical variability  or any
structural  variability  of  disk  and  circumstellar  materials.   We
compare  the NIR magnitudes  from 2MASS  \citep{cutri03} and  from the
more  recently obtained  JHK  magnitudes by  \citet{beltran09} of  the
sources  reported in both  the datasets  to identify  NIR variability.
Sources which have shown variation of $\sim$~0.1 mag in any of the NIR
band are listed in the last column of Table~\ref{tab_mirsrc}.

\subsection {Medium resolution spectra of YSOs
\label{sec_medspec}}

\begin{figure*}
\begin{center}
\resizebox{15cm}{!}{\includegraphics[angle=0,width=15cm,angle=0]{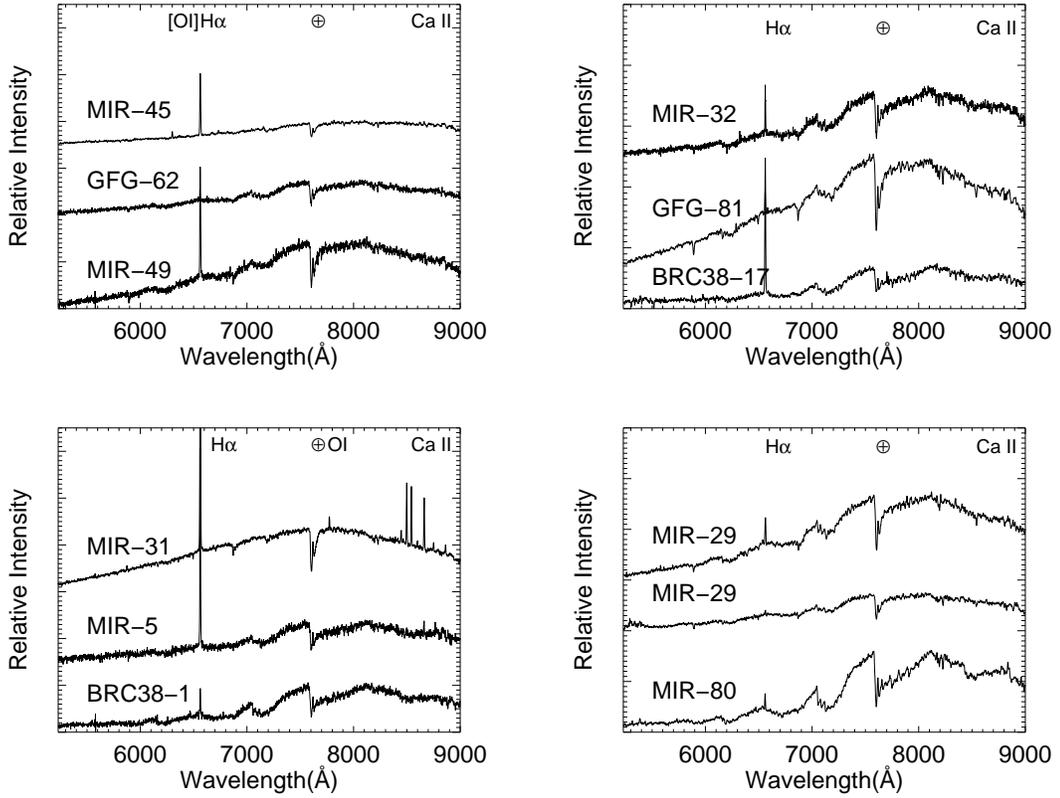}}
\caption{Sample spectra of YSOs in SFO 38 obtained with {\em HFOSC } instrument. 
The strong features e.g. \ha\ emission ($\lambda 6563$\AA\ ), CaII infrared triplet 
emission , OI line at $\lambda 7773$ \AA\ and [OI] line at $\lambda 6300$ \AA\ are
 marked in the spectra. Atmospheric features are indicated with $\earth$.
Two spectra of MIR-29 taken at two different epochs show
variable \ha\ emission (see text for details).}
\label{fig_specplot}
\end{center}
\end{figure*}

The \ha\ emission  though a very prominent feature  in the spectrum of
the YSO  is not sufficient to  identify YSOs because  there also exist
other kinds of \ha\ emission  line stars {\em viz.,} \textit{Be, Mira}
stars etc.   Medium resolution spectroscopy is  helpful to distinguish
the   YSOs    from   the   \textit{Be}    and   \textit{Mira}   stars.
Figure~\ref{fig_specplot} presents sample  spectra observed with HFOSC
of a few YSOs.  The prominent features e.g. \ha, CaII NIR triplet etc.
are  marked  in  the  respective  spectra.  We  have  detected  strong
\ha\ emission in  most of the Class II  objects except BRC\,38\,16 and
therefore  we do not  consider it  as a  YSO in  the remainder  of the
paper.  Log  of spectroscopic observations and  \ha\ equivalent widths
(W$_\lambda$)    of    the    individual    YSOs    are    given    in
Table~\ref{tab_sptclass}.  BRC~38~6 (MIR-31) is the brightest Class II
object detected  in optical  band.  From the  multiepoch spectroscopic
observations  we  find  that   MIR-31  shows  variable  \ha\  emission
(Table~\ref{tab_sptclass}).  We have  also detected OI $\lambda$7773 (
blend of OI $\lambda$7771.96, $\lambda$7774.18, $\lambda$7775.40 ), OI
$\lambda$8446 (blended with  Pa18 $\lambda$8437) CaII infrared triplet
i.e  $\lambda$8498 (blended  with  Pa16 $\lambda$8502),  $\lambda$8542
(blended with Pa15 $\lambda$8545) and $\lambda$8662 (blended with Pa13
$\lambda$8665) and  several other  Paschen lines e.g.   Pa10 $\lambda$
9014, Pa11 $\lambda$8862, Pa12 $\lambda$8750, Pa14 $\lambda$8598, Pa17
$\lambda$8467,  Pa19  $\lambda$8413,  Pa20  $\lambda$8392  in  MIR-31.
According  to   \cite{muzerolle98p1},  OI  $\lambda$7773   is  a  very
sensitive  indicator  of  infall  in  the  context  of  magnetospheric
accretion scenario.  LiI  $\lambda$6708 (W$_\lambda$ $\sim$0.7 \AA) is
also  present  in the  spectra  of  MIR-31.   All these  spectroscopic
features indicate that  MIR-31 is indeed a YSO  which is going through
an active  accretion phase.  We  have detected [OI]  $\lambda$6300 and
[OI]  $\lambda$6363 and  [SII] $\lambda$6717  and  [SII] $\lambda$6731
emission lines in  MIR-45 which are well known  signatures of outflows
from YSOs. We  also find a positive correlation  between \ha\ emission
and MIR excess  for YSOs with W$_\lambda$(\ha) $\le$~40  \AA.  We have
detected sporadic \ha\ emission  in transitional disk object MIR-29 on
3rd November 2008.  We observed the  object in 2009 also but we do not
find any  further \ha\ emission.  We have detected weak  \ha\ emission
from two  more transitional disk  objects e.g.  MIR-43  ([GFG2007] 62)
and MIR-76 ([GFG2007]  81).  We also obtained the  spectra of few more
transitional  disk  objects  e.g.   MIR-1, MIR-2,  MIR-9,  MIR-18  and
MIR-108 and do not find \ha\ emission in any of these objects.

\subsection {Spectral Classification}

We have  classified the observed optical spectra  by manual inspection
of  the spectra  of target  stars and  also by  comparing it  with the
spectra  of  known  spectral  type.  We observed  several  stars  with
spectral  type of  F to  M from  \cite{jacoby84},  \cite{valdes04} and
\cite{kirkpatrick91} with  HFOSC using the  same set up as  the target
stars  to   minimize  the   effect  of  instrumental   signatures  and
resolution.  From  preliminary analysis we  find that the YSOs  are of
spectral type K and M.

For more accurate  classification in the range of  spectral type K--M,
we  use   the  calibration  file   provided  in  the   {\em  SPTCLASS}
\citep{hernandez2004} code for various absorption features {\em viz.,}
Tio bands etc.  The  detailed information of the classification scheme
can       be       found       in       the       {\em       SPTCLASS}
code\footnote{\url{http://www.astro.lsa.umich.edu/~hernandj/SPTclass/
    sptclass.html}}.   This scheme measures  the equivalent  widths or
``indices" of the absorption features which are sensitive to T$_{eff}$
using  the adjacent  blue and  red continuum  bands and  calibrate the
indices against the spectral type of the stars.
Spectral type of the target  star is determined from the weighted mean
of the  spectral types obtained from different  absorption features of
the target  star excluding the  apparently anomalous values.   As this
classification  scheme  uses the  local  continuum  of the  absorption
feature, it is  insensitive to reddening and the  S/N ratio.  Further,
the  spectral features  which  are used  for  classification are  also
insensitive to the luminosity class.   The errors in the spectral type
determination come  from the measurement  of indices and  the standard
deviation  of the  spectral types  obtained from  different absorption
features.  To check the  applicability of  the adopted  calibration on
{\em HFOSC} spectra we classify the {\em HFOSC} spectra of a few K and
M type stars and retrieve  the spectral type within $\sim$0.5 subtype.
We then use  the same calibration scheme to  classify the YSO spectra.
We further  explore the Fe I  $\lambda$6495 \AA\ line  to classify the
spectra of  early K type  stars \citep{graybook09}.  The  ``index'' or
equivalent  width of  $\lambda$6495 line  was calculated  for spectral
types  F--K,  using the  {\em  HFOSC }  spectra  of  stars with  known
spectral type taken from  \cite{jacoby84,valdes04}. The results of our
classification  are  summarized  in Table~\ref{tab_sptclass}.  We
also tabulate  the effective temperatures  of YSOs using the  table of
\cite{kh95}.  Using the {\em V-I}  color of a particular spectral type
and  our  optical  {\em  BVRI}  observations we  calculated  the  \av\
 assuming the normal interstellar extinction law as given by
\cite{bessellir}.  The  average  extinction  towards  the  globule  is
estimated to be \av=~2.5 mag.


\subsection{Rate of accretion from \ha\ emission}

The \ha\ emission along with  the other Balmer lines, have always been
the  iconic   spectral  diagnostics  of   accretion  of  circumstellar
materials in  the YSOs. Magnetospheric accretion  models have provided
satisfactory explanation  of the observed  spectral characteristics of
accretion Classical  T Tauri Stars  (CTTS) e.g. broad  asymmetric line
profiles  etc.  \citep{muzerolle98model,muzerolle01p2}.   According to
the magnetospheric accretion model the stellar magnetic field disrupts
the circumstellar disk at several stellar radii and accreting material
fall  to   the  stellar  surface   along  the  magnetic   field  lines
\citep{koenigl91}.      Magnetospheric     accretion    models
  suggest that the hydrogen emission lines form in the infall
  zone,  so that the  blue shifted  asymmetric emission  line profiles
  arise due  to partial obscuration of  the flow by the  inner part of
  the accretion  disc and the  red shifted absorption  profiles result
  from infalling material at  near free-fall velocities on the stellar
  surface   \citep{muzerolle98model}.    Strength   of   the
\ha\ emission  line has been  widely used to distinguish  the actively
accreting  CTTS from their  less active  and evolved  counterparts i.e
Weak-lined T Tauri Stars  (WTTS).  \cite{white03} proposed to classify
a YSO  as CTTS if  W$_\lambda$ (\ha) $\geq$  3 \AA\ for  K0--K5 stars,
W$_\lambda$ (\ha) $\geq$ 10 \AA\ for K7--M2.5 stars, W$_\lambda$ (\ha)
$\geq$  20 \AA\  for M3-M5.5  stars, and  W$_\lambda$ (\ha)  $\geq$ 40
\AA\ for M6-M7.5  stars.  They also proposed that  the stars with full
width  of  \ha\ emission  line  at  10 $\%  $  of  the peak  intensity
(\ha[10$\%$])  greater  than  270   km  s$^{-1}$  are  accreting  CTTS
independent of spectral  type.  We identify {\em 9}  stars as CTTS and
{\em  5} stars  as  WTTS  according to  the  above mentioned  criteria
(Table~\ref{tab_sptclass}).   \cite{natta04}  showed that  \ha[10$\%$]
$\geq$ 200  km s$^{-1}$ can be  used to distinguish  the accretors and
non  accretors as well  as to  derive an  approximate estimate  of the
accretion rate \.{M}$_{ac}$  quantitatively by the following relation,
log\.{M}$_{ac}$~~=~~-12.89~($\pm$0.3)~+~9.7~($\pm$0.7)~\mult~10$^{-3}$~\ha[10$\%$]\\ where
\ha[10$\%$] is in km s$^{-1}$  and \.{M}$_{ac}$ is in \msun yr$^{-1}$.
In order to estimate the accretion rates for some of the well resolved
\ha\ line  profile we  fitted Gaussian profiles  to the  \ha\ emission
lines  and  calculate the  equivalent  width  and  FWHM of  each  line
profile.   The measured  FWHM  of each  profile  has been  deconvolved
assuming   a  Gaussian   instrumental  profile.    We   calculate  the
\ha[10$\%$]  for  these objects  and  transformed  them into  velocity
units. We  did not correct for the  underlying photospheric absorption
in  \ha\  if  there  is any.   Table~\ref{tab_sptclass}  presents  the
accretion rates \.{M}$_{ac}$, calculated using the above formula along
with  the  measured values  of  \ha[10$\%$].   Based  on the  observed
multiepoch  spectra we  find variable  accretion  rates in  BRC 38  1,
MIR-5, MIR-31, MIR-32 and MIR-45.

\subsection{Optical Color-Magnitude Diagram}

\begin{figure*}[ht]
\begin{center}
\resizebox{\hsize}{!}{\includegraphics[angle=0,width=18cm,angle=0]{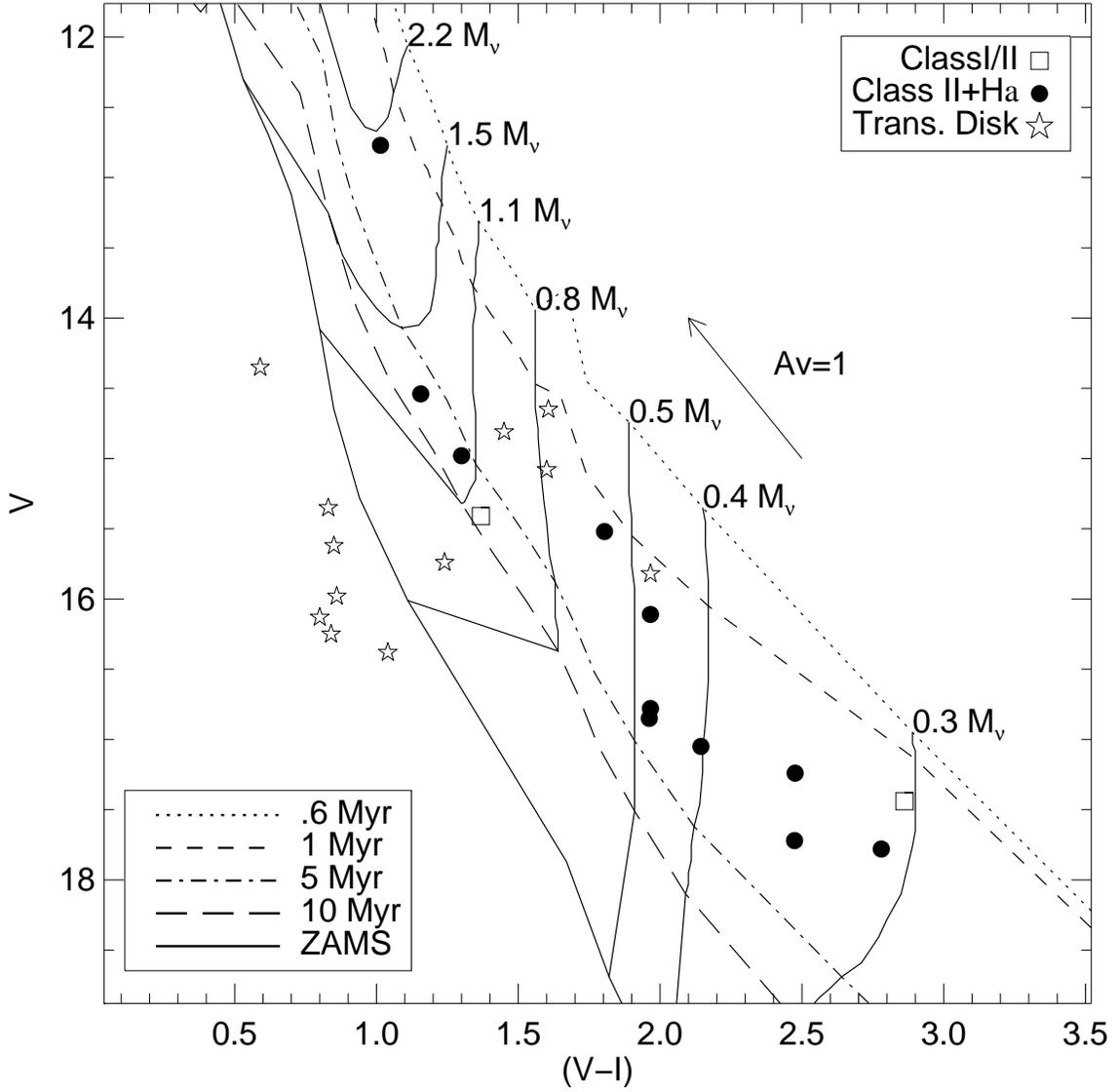}}
\caption{Extinction corrected V vs. V-I Color-Magnitude diagram of YSOs near SFO 38. \ha\ emission YSOs are shown by {\em filled circles}. The ZAMS and 0.6, 1, 5, 10 Myr 
isochrones of \cite{siesspms} are shown with the evolutionary tracks
for masses from 0.3 to 2.2 \msun, at an adopted distance of 750~pc.
\label{fig_vicm}}
\end{center}
\end{figure*}

Figure~\ref{fig_vicm}  presents the  extinction corrected  V  vs.  V-I
diagram  of  the  YSOs  near  SFO\,38.   We  have  corrected  for  the
extinction of individual objects with known spectral types and used an
average  value of \av=2.5  mag when the  spectral type  is not  known.  In
order  to estimate  the  mass and  age  of the  YSOs  we overplot  the
isochrones  from   \cite{siesspms}  assuming  a   distance  of  750~pc
\cite{matthews1979}. We find that most of the YSOs (Class II with(out)
\ha) have ages between 1-5 Myr  and masses in the range 0.3-2.2 \msun.
We note that several  transitional disk objects are located to
  the left of  the ZAMS line. A closer examination  of the position of
  the  transitional  disk  objects  in  the  MIR  color-color  diagram
  (Fig.~\ref{fig_iracmipscol}) reveals  that 6 out of  7 transitional disk
  objects which are  situated to the left of  ZAMS line have [5.8]-[8]
  color excess $<$ 0.5 and 4  out of 5 transitional disk objects which
  are situated to  the right of ZAMS line  have [5.8]-[8] color excess
  $>$ 0.5.  The transitional disk objects to the left of the ZAMS line
  are  more  like Class  III  objects  with  smaller values  of  color
  excess.  We  have  used   an  average  \av=2.5  mag  for  extinction
  correction  for these  objects because  the spectral  types  are not
  known. There  are several factors  that could lead  the transitional
  disk objects  to the  anomalous position in  Fig.~\ref{fig_vicm} e.g.
  extinction correction,  uncertainties in distance  and PMS evolution
  model used.   An extinction correction with a  smaller \av\ ($\leq$1
  mag) brings all  the transitional disk objects closer  to, or to the
  right of the ZAMS line.  The  distance of IC~1396 is in the range of
  615~pc \citep{dezeeuw99} to  900~pc \citep{contreras2002} and we adopt
  a distance  of 750~pc for SFO~38  to make our  results comparable to
  most   published  results.   This  uncertainty   in   the  distance
  corresponds  to an  uncertainty of  $\pm$~0.4 mag  in  the absolute
  magnitude.   Adoption  of  a   distance  of  900~pc  brings  up  the
  transitional disk objects somewhat closer to the ZAMS line but these
  objects still occupy positions  below the ZAMS line. While different
  theoretical models differ in the details of the PMS evolution, close
  to the ZAMS line all these models agree with each other and with the
  empirical ZAMS of the  young star clusters. Therefore, the anomalous
  positions of the transitional disk  objects with respect to the ZAMS
  track in  Fig.~\ref{fig_vicm} is primarily due  to overcorrection of
  the   extinction.   Estimation   of  mass   and   age  from   the
color-magnitude  diagram (CMD)  is strongly  dependent on  the adopted
model \citep{hillenbrand2008}.  The  ages can differ by as  much as 10
Myr or  more when two  different models are  used on the  same dataset
\citep{ikeda08}.  The main  source of error in the  V vs.  V-I diagram
is  the estimation  of extinction  towards the  globule.  In  the last
column of Table~\ref{tab_sptclass} we list the ratios of E(V-I)/E(B-V)
which are different  from the standard value of  1.6 and indicate that
normal extinction law  may not hold good for  star forming regions due
to plenty of  patchy obscuring materials.  Although the  NIR {\em JHK}
bands suffer comparatively less extinction than the optical {\em BVRI}
band but for the YSOs,  thermal emission also contributes to {\em JHK}
bands  resulting  in  excess  fluxes  compared to  the  main  sequence
counterparts without disk.  For a  given model we consider the derived
values of  the masses  and ages as  representative values and  the age
spread of YSOs as more reliable estimate.

\subsection{SED of YSOs with MIPS 24~\micron\ detection}

In  order to  interpret the  observed SEDs  for the  YSOs  detected at
24~\micron\ and better characterize these sources we have explored the
archive of two-dimensional (2D) axisymmetric radiative transfer models
of  protostars calculated for  a large  range of  protostellar masses,
accretion  rates,  disk  masses   and  disk  orientations  created  by
\citet{robitaille2007}.    This  archive   also   provides  a   linear
regression tool which can select  all model SEDs that fit the observed
SED better than a specified  $\chi^2$.  Each SED is characterized by a
set of model  parameters, such as stellar mass,  temperature, and age,
envelope  accretion rate, disk  mass, and  envelope inner  radius.  We
have used this online tool  to generate models, which fit the observed
SEDs  for the  14 YSOs  which have  been detected  at  24~\micron.  We
restricted the SED fitting tool  to explore only distances between 650
and 850~pc. Below we present and briefly discuss the best fit models.

\begin{figure*}
\begin{center}
\resizebox{\hsize}{!}{\includegraphics{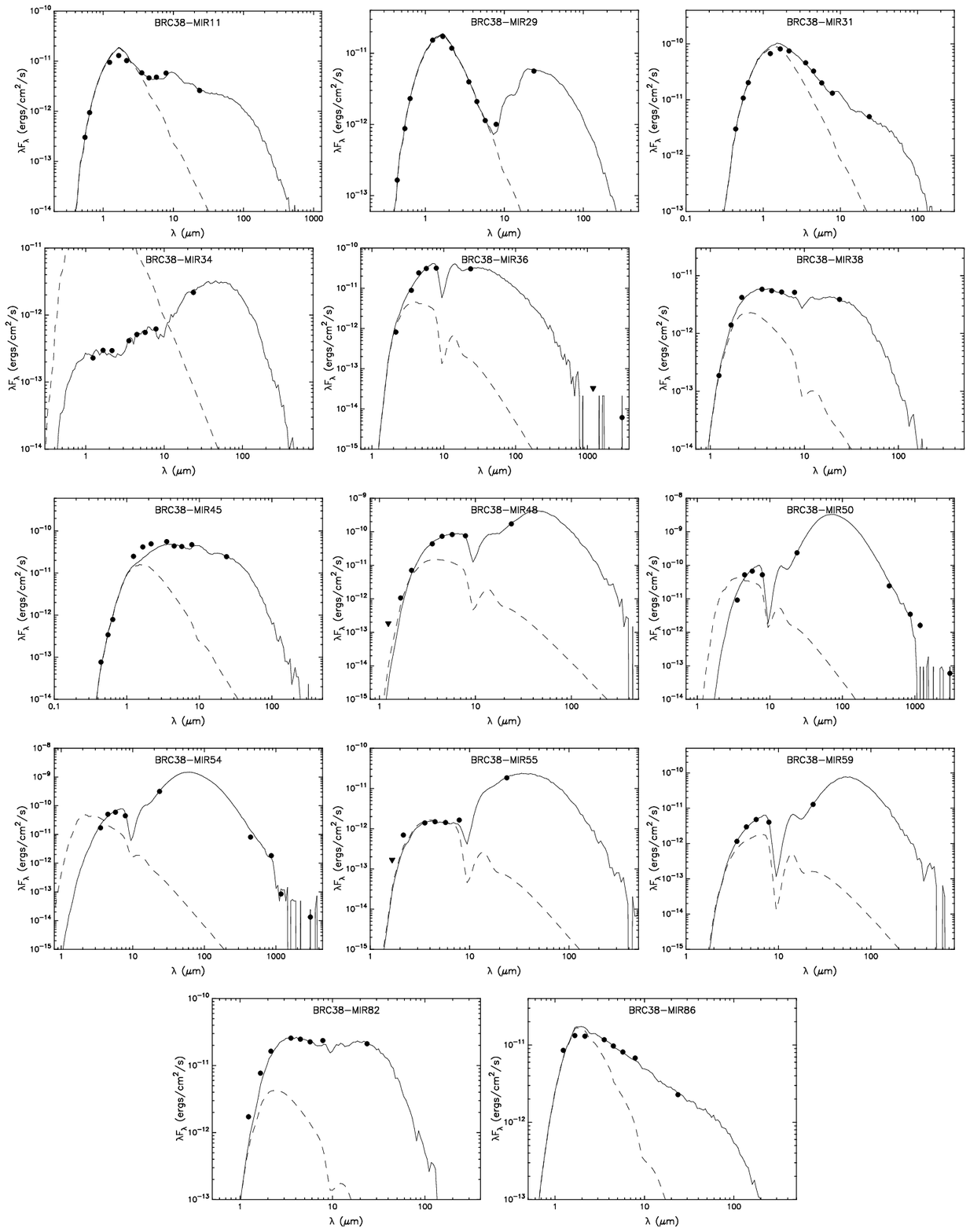}}
\caption{SED  fits for  the PMS  objects using  axisymmetric radiation
  transfer models.   The filled  circles indicate the  measured fluxes
  and  uncertainties. The  filled  triangles correspond  to the  upper
  limits  of flux  at these  wavelengths.  The  continuous  black line
  represents  the  best-fitting SED, 
  and the  dashed line  shows the
  stellar photosphere corresponding to  the central source of the best
  fitting model,  in the absence of circumstellar  dust (but including
  interstellar extinction).}
\label{fig_sedfits}
\end{center}
\end{figure*}

Figure~\ref{fig_sedfits}  shows results  of detailed  modeling  of the
observed SED in  the mid-infrared (and sub-mm for  MIR-50) for all the
YSOs which were detected at 24~\micron.  A major criticism against the
use  of these  models has  been  the non-uniqueness  of the  solutions
obtained from the model library.  However we find that for all sources
the best  fit models are  distinctively better than the next few
models in reproducing  all the observed   flux   densities.
Table~\ref{tab_sedfits} presents  the parameters corresponding  to the
best  fit  models  for  the  fourteen sources  which  we  fitted  with
the accretion disk models.

\citet{robitaille2006}  presented  a  classification scheme  which  is
essentially analogous  to the Class  scheme, but refers to  the actual
evolutionary stage of the object based on its physical properties like
disk mass  and envelope accretion rates  rather than the  slope of its
near/mid-IR   SED.   According   to   the  \citeauthor{robitaille2006}
classification scheme, MIR-50 and MIR-54, are stage 0/I objects, while
all the  others are stage II  objects.  However, the  YSOs MIR-34, 36,
48,  55 and  59 were  identified  as Class  0/I objects  based on  the
color-color diagrams. Next we consider the fact that except for MIR-34
all the  sources have rather  high visual extinction (\av\  $>$40) and
the MIR color-color diagrams we had  used to classify the YSOs did not
correct for the reddening. Using standard interstellar extinction laws
\citep{indebetouw05}and the \av\ derived from the models, we find that
when  de-reddened, the  sources MIR-48,  55 and  59 occupy  the region
typically occupied by Class II objects.  However for MIR-34 and 36 the
difference  in the inferred  evolutionary stage  can not  be explained
only in terms of reddening.

Based  on  the results  of  fitting  the  observed SEDs  with  axially
symmetric accretion  models we  find that all  the YSOs  analyzed here
have low to  intermediate-mass central stars. Figure~\ref{fig_ysomap}
shows that consistent with our expectations the sources located within
the  globule  have \av  $>$20~mag,  while  the  sources lying  at  the
periphery  of   the  bright  rim  have  \av\   between  3--5~mag.  In
Sec.~\ref{sec_mir5054}  we discuss  the  two most  luminous Class  0/I
sources MIR-50 and MIR-54 in detail.

The ages estimated for these fourteen YSOs from the SED models and the
location  of these sources  in BRC~38  show that  most of  the sources
closer  to the  bright rim  have ages  between 5--8~Myr.   The sources
within the  BRC and lying almost along  the axis of the  BRC have ages
between 0.1--1~Myr,  with the protostellar cluster at  the position of
IRAS 21391+5802 being  the youngest with an age  of $\sim$0.1~Myr. The
age spread  estimated for  the YSOs (mostly  Class II and  above) from
optical  CMD diagram  is of  the range  1--8 Myr.   However,  the ages
estimated from both methods are higher than the ages obtained from the
NIR CMD by \cite{getman07}.

\subsection{The protostellar cluster at IRAS 21391+5802
\label{sec_mir5054}}

Interferometric millimeter  continuum images have shown  that the most
luminous source IRAS 21391+5802 in  the IC~1396 region can be resolved
into   three  components,   named   as  BIMA~1,   BIMA~2  and   BIMA~3
\citep{beltran02}. Radio continuum images  at 3.6~cm observed with VLA
have  also  identified  three  peaks  VLA~1--3,  coincident  with  the
millimeter continuum  peaks \citep{beltran02}. Of  these three sources
VLA~3   is  the   strongest  emitter   of  radio   continuum.   Recent
sub-arcsecond resolution millimeter observations have identified three
components within  BIMA~2 itself  \cite{neri07}.  The SCUBA  images at
450  and  850~\micron\ do  not  resolve  the  components of  continuum
emission associated with IRAS 21391+5802 \citep{morgan2008}.

We identify MIR-50  and 54 as the mid-infrared  counterparts of BIMA~2
and BIMA~3  and do not detect  any source associated  with BIMA~1. The
source MIR-36 is  located 3\farcs8 away from BIMA~1  and hence is most
likely not associated with BIMA~1.   We have used the continuum fluxes
available in literature to construct  the observed SEDs for MIR-50 and
MIR-54. In order to make a realistic estimate of the flux densities at
450 and 850~\micron\ from each of  BIMA~2 and BIMA~3, we have split up
the observed  total flux densities  at these wavelengths  according to
the ratio  of flux  densities of these  sources measured at  1.2~mm by
\citet{beltran02}.  Fitting of  the observed  flux densities  with the
axially symmetric  accretion-based models \citep{robitaille2007} shows
that   MIR-50  (BIMA~2)   is  a   protostellar  object   with   $L$  =
197~\lsun\ and a  mass of 5.97~\msun. Both mass  and luminosity values
match reasonably  well with the  literature. We note however  that the
model fitted does not reproduce the interferometric flux densities at
1.2  and 3.1~mm  well.  This  is consistent  with  the conclusions  of
\citet{beltran02}  that   the  interferometric  observations   miss  a
significant  amount  of  emission.  The  best  fit  model  for  MIR-54
corresponds to  a mass of  1.5~\msun\ and a luminosity  of 33.4~\lsun.
\citet{codella01}   had   estimated   an  integrated   luminosity   of
233~\lsun\ including both BIMA~2 and BIMA~3 and here we derive a total
luminosity of  230~\lsun. Based on  the SED fitting performed  in this
paper, MIR-54 or  BIMA~3 is a low mass Class  0/I protostar and MIR-50
or   BIMA~2  is   a   Class  0/I   intermediate-mass  star,   although
\citet{neri07}   have   already   shown   BIMA~2   as   having   three
components. The SED models derive the  age of both these sources to be
$\sim 10^5$~yr.

\section{Discussion: RDI mechanism at work in SFO 38
\label{sec_dis}}

Based  on  the  observational  evidences  in  the  form  of  YSOs,  an
outflow-driving  IRAS  source  and  several  HH  objects,  SFO\,38  is
unambiguously  an  active  star   forming  region.  There  is  however
considerable  disagreement  regarding whether  the  star formation  is
triggered or  influenced by  the nearby OB  association. The  two most
recent studies  of SFO\,38  based on Chandra  X-ray and  {\em Spitzer}
Mid-IR observations \citep[up to 5.8~\micron][]{getman07} and deep NIR
observations   \citep{beltran09}  arrive  at   diametrically  opposite
conclusions.  Considering the spatial and temporal distribution of the
fewer YSOs selected by using MID-IR color-color diagram ( [3.6]--[4.5]
vs.  [4.5]--[5.8] ), \cite{getman07} argued in favor of triggered star
formation  by   Radiation  Driven  Implosion   (RDI)  model.   However
\cite{beltran09} did not find either {\em (a)} enhanced star formation
activity towards the  southern rim of SFO\,38 or  {\em (b)} systematic
gradient  in the evolutionary  stage of  the YSOs  using the  deep NIR
imaging.  Based   on  the  absence   of  NIR  excess  from   the  YSOs
\citet{beltran09} concluded that the sources closer to the O type star
appear to be  more evolved primarily due to  the modification of their
environment   by   the   intense   UV  field.   Thus,   according   to
\citeauthor{beltran09}   the   evolutionary   sequence   observed   by
\citeauthor{getman07} should not necessarily  be interpreted as due to
triggered star formation.  However \citeauthor{beltran09} also concede
that NIR-only studies  are inadequate to identify all  the YSOs in the
region.

We explore two  possible scenarios which could have  been the dominant
driving mechanism  to initiate the formation  of stars in  SFO 38 {\em
  viz.,}  coeval  star  formation  throughout  the  \HII\  region  and
triggered star formation by the RDI mechanism. The assumption that the
stars near the globule were formed simultaneously with the other stars
of the OB association suggests that the distribution of YSOs should be
correlated with the  distribution pattern of the other  massive to low
mass stars of the  OB association. The observed elongated distribution
of YSOs  along two different axes  in SFO\,38 is  quite different from
the  distribution  of  the   known  members  of  Cep  OB2  association
\citep[Fig.  22 of][]{dezeeuw99}.   Moreover the  ages of  the outflow
driving  IRAS  source  and  the associated  protostellar  cluster  are
significantly less than the dynamical  age of 3 Myr \citep{patel95} of
the \HII\  region created by the  massive OB type stars.  We thus rule
out  the possibility  of  coeval star  formation  in SFO  38 with  the
members of Cep OB2 association, although some of the \halpha\ emitting
sources  could have  been formed  prior to  the arrival  of  the shock
front. Next,  we investigate the  other possibility, based on  the RDI
model, by comparing our  observations with the predicted signatures of
the model.  The RDI  model suggests enhancement  of density  along the
axis of the UV irradiated globule and an age sequence of YSOs with the
oldest YSO lying closer to the OB association.

Sequential  star  formation  is  considered  to  be  one  of  the  key
signatures of triggered star formation via RDI. \citet{beltran09} have
cautioned against  the use  of the spatial  distribution of YSOs  as a
proof of triggered  star formation, by arguing that  intense UV fields
close enough to the OB associations tend to destroy the protoplanetary
disks around the  YSOs and thus make them appear  to be older. However
we find  this argument  untenable for the  case of SFO\,38,  since the
typical  radius  of influence  of  O stars  is  found  to be  $<1$\,pc
\citep{balog2007,hernandez2008,mercer2009},  while the  mean projected
distance  of the  nearest O  and  B -type  stars close  to SFO\,38  is
$\sim$9\,pc.  Thus  the spatial distribution of the  YSOs discussed in
SFO\,38 signify the temporal evolution of  the YSOs and not due to the
destruction of  protoplanetary atmospheres under the  influence of the
UV radiation from the OB association.

The bright-rimmed cloud SFO\,37, was found to have an an exemplary bow
shaped  geometry  and  show  a  clean evolutionary  sequence  of  YSOs
\citep{ikeda08}. SFO\,38 in contrast  has a more complicated geometry.
As noted earlier the \halpha\ emission from SFO\,38 shows an asymmetry
relative  to the axis  connecting the  globule to  the nearest  O star
HD\,206267. Further,  in SFO\,38 though  the YSOs clearly show  an age
segregation, they are not distributed along a single straight line. It
is however possible to identify  another direction along which some of
the YSOs are  most likely aligned. The asymmetry  of \halpha\ emission
and the  non-linearity of  the distribution of  YSOs suggest  that the
evolution of SFO\,38 is influenced  by the nearby B star HD\,206773 as
well.  From the  \ha\ line  image (Figure~\ref{fig_ysomap})  HD 206773
seems to be ionizing the eastern part of the rim more efficiently than
the extent to which HD 206267 has ionized the western part of the rim.
Using the $^{13}CO$~ {\em J}~=~1--0 map of SFO~38, \cite{patel95} also
found that the eastern and  western wings of SFO~38 differ in velocity
by about 1.5 km~$s^{-1}$ relative to the velocity of the dense core of
the globule. This suggests that HD\,206773 plays an active role in the
present dynamics of the globule.

We have identified a  spatio-temporal gradient along the directions of
both HD\,206267  and HD\,206773.  However the spatial  gradient is not
continuous, there is a distinguishable  gap between the two classes of
YSOs  e.g.  the projected  distance  of  Class  I/II sources  from  the
approximate center  of the distribution  of Class I sources  is $\sim$
0.3~pc and this gap can be interpreted as due to the difference in the
arrival times of the shock within  the framework of the RDI model.  If
we  consider the  age spread  to be  $\sim$ a  few Myr  (comparing the
estimated  ages from  SED fitting  and  optical CMD)  and the  spatial
separation between the MIR-50 (Class  0/I) and MIR-29 (Class II) to be
$\sim$  0.6~pc,  we  obtain  the  speed of  shock  propagation  to  be
0.1--0.3\,km\,$s^{-1}$. The  estimated shock speed  is consistent with
the   shock  velocities  obtained   from  3-D   numerical  simulations
\citep{miao06}.   The  estimated  speed   of  shock   propagation  are
approximate since we use  only projected distances between sources and
do not  consider the time lapse  between the arrival  of the shockwave
and  the triggering  of  star  formation in  the  globule, which  from
simulations is $\sim$ 0.375~Myr \citep{miao06}.

Thus,  our observations  and subsequent  estimate of  the evolutionary
stages of the YSOs embedded in SFO\,38, are consistent with a scenario
of  triggered star  formation via  the  RDI mechanism  powered by  two
different ionizing sources, HD\,206267 and HD\,206773.

\section{Summary 
\label{sec_conclusion}}

We  have presented  multiwavelength photometric  study of  SFO~38 from
optical {\em BVRI}  to {\em Spitzer} IRAC and  MIPS observations along
with optical  spectroscopy of the  selected objects. We  summarize our
work as follows.

\begin{itemize}
 \item  A  total  of  40  YSOs (Classes  0/I/II)  and  13  YSO
   candidates are  identified based on  MIR color indices and  we also
   confirm \ha\  emission from  2 YSOs which  are not known  in the
   literature.  We  further  identify   4  additional  YSOs  based  on
   \ha\ emission and thus we find 44 YSOs in and around SFO~38.

 \item \ha\ emission line YSOs are  mostly K--M type stars with an age
   spread of  1-8 Myr and mass range  of .3-2.2 \msun\.  Some of the
   YSOs show  photometric variation in  optical and NIR  bands and
   also variable \ha\ emission  in the medium resolution spectra. Mass
   accretion rates estimated form broad  \ha\ line profiles are of the
   order of 10$^{-8}$ to 10 $^{-10}$ \msun\ yr$^{-1}$.

 \item Mass, luminosity  and age for the different  components of the
   protostellar cluster at IRAS 21391+5802 were derived by SED fitting.

 \item   Continuum  subtracted   \ha\  line   image   show  asymmetric
   \ha\  emission at  the  bright-rim. Two   OB  type
     stars e.g.  HD  206267 (O6.5 ) and HD  206773 (B0V) are proposed
   as the potential ionizing sources.

 \item Class II  to Class 0/I objects are distributed  from the rim to
   the  core part  of  the  globule respectively.  It  is possible  to
   identify  at least  two different  axes  of elongation  of the  YSO
   distribution. The spatial gap  between the two different classes of
   YSOs is  consistent with the difference  in the arrival  time of a
   shockwave  propapagating  into  the globule.   The  spatio-temporal
   gradient in the distribution of  YSOs along two different axes that
   are parallel to one of  the either ionizing star indicate triggered
   star formation due to the Radiation Driven Implosion model.
        
\end{itemize}

This work  is based  [in part] on  observations made with  the Spitzer
Space Telescope,  which is operated by the  Jet Propulsion Laboratory,
California Institute  of Technology under a contract  with NASA.  This
research  has made  use  of the  SIMBAD  data base,  operated at  CDS,
Strasbourg, France.   This publication makes use of  the data products
from  the  2MASS,  which is  a  joint  project  of the  University  of
Massachusetts    and   the    Infrared    Processing   and    Analysis
Center/California  Institute  of Technology,  funded  by the  National
Aeronautics  and   Space  Administration  and   the  National  Science
Foundation.  RC would  like to thank Dr.  P. S.   Parihar and Ramya S.
for useful discussions.

\newpage
\setcounter{table}{0}
\begin{deluxetable}{lcclllrrrrrrl}
\tabletypesize{\tiny}
\rotate
\tablecaption{Results of Near- \& Mid-IR photometry of the Mid-infrared sources in SFO 38
\label{tab_mirsrc}}
\tablewidth{0pt}
\tablehead{
\colhead{Source} &
\colhead{$\alpha_{2000}$}&
\colhead{$\delta_{2000}$}&
\colhead{$J$}&
\colhead{$H$}&
\colhead{$K_{\rm s}$}&
\colhead{$F_{3.6}$}&
\colhead{$F_{4.5}$}&
\colhead{$F_{5.8}$}&
\colhead{$F_{8.0}$} &
\colhead{$F_{24}$}& 
\colhead{Class} &
\colhead{Comments}\\
SFO38 & &&&&&mJy&mJy&mJy&mJy&mJy&\\}
\startdata
MIR-1 & 21:40:24.97 &  58:16:10.5 &      14.48$\pm$0.04  &    13.72$\pm$ 0.05  &
13.42$\pm$ 0.05 &     1.67$\pm$0.01&     1.10$\pm$0.01& 0.96$\pm$0.02&
1.03$\pm$0.02&  \ldots & Trans. & {\em $JHK_s$}\tablenotemark{c}\\
MIR-2 & 21:40:25.39 &  58:15:47.3 &  15.35$\pm$0.06  &    14.51$\pm$ 0.07  &
14.15$\pm$ 0.07 &     0.94$\pm$0.01&     0.63$\pm$0.01& 0.68$\pm$0.02&
0.98$\pm$0.02&  \ldots & Trans. & \ldots\\
MIR-3 & 21:40:25.84 &  58:15:26.1 &   15.62$\pm$0.07  &    14.80$\pm$ 0.07  &    14.62$\pm$ 0.10 &     0.52$\pm$0.01&     0.31$\pm$0.00&
0.23$\pm$0.02&     0.07$\pm$0.02&  \ldots &III/S & \ldots\\
MIR-4 & 21:40:26.84 &  58:14:51.8  &   16.26$\pm$0.10  &    15.67$\pm$ 0.15  &    15.06$\pm$ 0.15 &     0.32$\pm$0.00&     0.20$\pm$0.00&
0.17$\pm$0.02&     0.11$\pm$0.02&  \ldots &III/S  & \ldots\\
MIR-5 & 21:40:27.29 &  58:14:21.14 & 14.30$\pm$0.04  &    13.30$\pm$ 0.04  &    12.88$\pm$ 0.04 &     6.22$\pm$0.05& \ldots& \ldots&
\ldots& \ldots & II & \ha, X-ray\\
MIR-6 & 21:40:27.38 &  58:16:21.3 &      \ldots  &     \ldots  &     \ldots &     0.07$\pm$0.00&     0.03$\pm$0.00&     0.51$\pm$0.02&
2.00$\pm$0.03&  \ldots &I/II\\
MIR-7 & 21:40:27.97 &  58:15:14.1 &    14.51$\pm$0.04  &    13.41$\pm$ 0.04  &
12.94$\pm$ 0.04 &     3.89$\pm$0.02&     3.52$\pm$0.02& 3.44$\pm$0.03&
4.33$\pm$0.03&  \ldots & II & \ha, {\em J}\\
MIR-8 & 21:40:28.35 &  58:16:42.3 &     11.59$\pm$0.03  &    10.78$\pm$ 0.03  &    10.43$\pm$ 0.02 &    23.62$\pm$0.07&    14.18$\pm$0.04&
10.94$\pm$0.06&     6.63$\pm$0.04&  \ldots & III/S\\
MIR-9 & 21:40:30.63 &  58:15:00.0 &     15.31$\pm$0.05  &    14.90$\pm$ 0.07  &    14.70$\pm$ 0.11 &     0.49$\pm$0.01&     0.32$\pm$0.01&
0.28$\pm$0.01&     0.22$\pm$0.02&  \ldots & Trans.\\
MIR-10 & 21:40:31.00 &  58:15:09.2 &    13.44$\pm$0.03  &    12.97$\pm$ 0.04  &    12.79$\pm$ 0.04 &     2.61$\pm$0.02&     1.63$\pm$0.01&
1.29$\pm$0.02& \ldots&  \ldots & III/S\\
MIR-11 & 21:40:31.56 &  58:17:55.1 &     14.03$\pm$0.03  &    12.89$\pm$ 0.03  &    12.39$\pm$ 0.03 &     6.93$\pm$0.03&     6.93$\pm$0.03&
9.15$\pm$0.06&    15.17$\pm$0.12&     20.46$\pm$0.08 & II & \ha, X-ray, {\em $JHK_s$}\\
MIR-12 & 21:40:31.97 &  58:14:56.0 &    16.72$\pm$0.16
&    15.78$\pm$ 0.15  &    15.41$\pm$ 0.21 &     0.41$\pm$0.00&
0.28$\pm$0.00&     0.26$\pm$0.01& \ldots&  \ldots &III/S \\
MIR-13 & 21:40:32.23 &  58:16:53.6 &    14.13$\pm$0.03  &    12.44$\pm$ 0.04  &    11.71$\pm$ 0.03 &    10.82$\pm$0.04&     6.39$\pm$0.02& 6.54$\pm$0.05&     7.36$\pm$0.05&  \ldots & Trans.\\
MIR-14 & 21:40:32.46 &  58:13:47.2 &     13.40$\pm$0.03  &    12.83$\pm$ 0.03  &    12.63$\pm$ 0.03 &     2.78$\pm$0.03& \ldots&
1.30$\pm$0.05& \ldots&  \ldots& III/S \\
MIR-15 & 21:40:32.71 &  58:17:23.5 &     12.19$\pm$0.03  &    11.55$\pm$ 0.03  &    11.36$\pm$ 0.03 &     9.21$\pm$0.04&     5.50$\pm$0.02&
4.04$\pm$0.04&     1.71$\pm$0.04&  \ldots& III/S \\
MIR-16\tablenotemark{b} & 21:40:32.84 &  58:15:47.8 & 17.42$\pm$0.01& 16.61$\pm$0.01 & 16.09$\pm$0.01 &     1.42$\pm$0.01&     0.73$\pm$0.01&     9.50$\pm$0.05& 32.10$\pm$0.11&  \ldots & I/II\\
MIR-17 & 21:40:33.64 &  58:14:44.6 &  16.54$\pm$0.14  &    15.81$\pm$ 0.18  &    15.07$\pm$ 0.00 &     0.21$\pm$0.00&     0.13$\pm$0.00&
0.11$\pm$0.01&     0.16$\pm$0.02&  \ldots & Galaxy\\
MIR-18 & 21:40:33.69 &  58:14:59.5 &     13.06$\pm$0.03  &    12.09$\pm$ 0.03  &    11.75$\pm$ 0.02 &     8.13$\pm$0.03&     4.82$\pm$0.02&
4.72$\pm$0.04&     5.60$\pm$0.03&  \ldots & Trans.\\
MIR-19 & 21:40:34.05 &  58:18:08.1 &  15.70$\pm$0.06  &    13.40$\pm$ 0.04  &    12.30$\pm$ 0.02 &     7.41$\pm$0.03&     5.16$\pm$0.02&
4.95$\pm$0.04&     4.03$\pm$0.04&  \ldots & Trans.\\
MIR-20 & 21:40:34.30 &  58:18:23.6 &  \ldots  &     \ldots  & \ldots &     0.31$\pm$0.00&     0.84$\pm$0.01&     1.03$\pm$0.02&
1.24$\pm$0.04&  \ldots& H$_2$ \\
MIR-21 & 21:40:34.59 &  58:16:11.4 &  \ldots  &     \ldots  &     \ldots &     0.08$\pm$0.00&     0.05$\pm$0.00&     1.33$\pm$0.03&
\ldots&  \ldots & III/S\\
MIR-22\tablenotemark{b} & 21:40:34.76 &  58:15:20.9 &   17.70$\pm$0.01 &  16.68$\pm$0.01 &  16.04$\pm$0.01 &     0.25$\pm$0.01&     0.34$\pm$0.01& \ldots& \ldots&  \ldots&
III/S \\
MIR-23 & 21:40:34.79 &  58:16:10.8 &  \ldots  & \ldots  &     \ldots & \ldots& \ldots& \ldots&     4.47$\pm$0.05& \ldots& \ldots \\
MIR-24 & 21:40:35.02 &  58:18:22.2 &   16.48$\pm$\ldots & 15.72$\pm$\ldots  & 14.90$\pm$ 0.21  &         0.62$\pm$0.01& 2.68$\pm$0.01&
1.84$\pm$0.03&     1.42$\pm$0.03&  \ldots & H$_2$ \\
MIR-25\tablenotemark{b} & 21:40:35.11 &  58:15:30.9 &   18.21$\pm$0.02
&    16.85$\pm$0.01 & 16.16$\pm$0.02 &     0.92$\pm$0.01&     0.48$\pm$0.01&     5.32$\pm$0.04& 18.26$\pm$0.08&  \ldots &I/II \\
MIR-26 & 21:40:35.20 &  58:14:05.0 &     13.11$\pm$0.03  &    12.61$\pm$ 0.03  &    12.45$\pm$ 0.03 &     3.55$\pm$0.02&     2.40$\pm$0.01&
1.59$\pm$0.02&     0.84$\pm$0.02&  \ldots & III/S \\
MIR-27 & 21:40:35.68 &  58:18:21.3 &    16.55$\pm$\ldots & 15.73$\pm$\ldots  & 14.79$\pm$ 0.17  &    0.77$\pm$0.01& 3.61$\pm$0.02&
2.20$\pm$0.03&     1.91$\pm$0.03&  \ldots & H$_2$\\
MIR-28 & 21:40:36.46 &  58:15:23.6 &   15.97$\pm$0.10  &    14.73$\pm$ 0.08  &    14.04$\pm$ 0.07 &     2.27$\pm$0.01&     1.43$\pm$0.01&
6.68$\pm$0.04&    19.47$\pm$0.08&  \ldots & I/II\\
MIR-29 & 21:40:36.52 &  58:13:45.8 &  13.51$\pm$0.02  &    12.58$\pm$ 0.03  &    12.24$\pm$ 0.03 &     4.69$\pm$0.02&     3.14$\pm$0.01&
2.17$\pm$0.03&     2.64$\pm$0.03&     44.29$\pm$0.09  & Trans. & \ha, X-ray\\
MIR-30 & 21:40:36.65 &  58:18:23.0 &    17.12$\pm$\ldots & 15.18$\pm$0.10 & 13.93$\pm$0.07 &     1.84$\pm$0.01& 1.55$\pm$0.01&
1.42$\pm$0.03&     1.21$\pm$0.03&  \ldots & II \\
MIR-31 & 21:40:36.88 &  58:14:37.8 &    11.90$\pm$0.02  &    10.89$\pm$ 0.03  &    10.23$\pm$ 0.02 &    54.30$\pm$0.14&    48.79$\pm$0.10&
38.25$\pm$0.12&    34.45$\pm$0.10&     39.36$\pm$0.07& II  & \ha, X-ray\\
MIR-32 & 21:40:37.02 &  58:15:02.9 &     14.27$\pm$0.03  &    13.28$\pm$ 0.04  &    12.82$\pm$ 0.03 &     4.01$\pm$0.02&     3.07$\pm$0.01&
2.86$\pm$0.03&     3.87$\pm$0.04&  \ldots& II  & \ha, {\em $K_s$}\\
MIR-33 & 21:40:37.16 &  58:15:46.6 &  17.01$\pm$\ldots & 15.80$\pm$0.16 & 14.55$\pm$0.10  &    1.33$\pm$0.01&     1.28$\pm$0.01&
0.71$\pm$0.03& \ldots&  \ldots & III/S\\
MIR-34\tablenotemark{b} & 21:40:37.54 &  58:15:39.3 &   18.07$\pm$0.02 &    16.98$\pm$0.01 &  16.24$\pm$0.02 &     0.49$\pm$0.01&     0.77$\pm$0.01&     1.05$\pm$0.03& 1.63$\pm$0.04&     17.15$\pm$0.06 & 0/I\\
MIR-35 & 21:40:39.05 &  58:16:03.3 &    \ldots  &     \ldots  & \ldots &     0.41$\pm$0.01&     1.04$\pm$0.01&     2.19$\pm$0.03& \ldots&
\ldots & H$_2$ \\
MIR-36\tablenotemark{a} & 21:40:39.60 &  58:16:09.0    & \ldots & $>$17.6 &
15.14$\pm$0.16  &     10.47$\pm$0.03&    36.11$\pm$0.08&    59.09$\pm$0.17&
82.89$\pm$0.22&    239.10$\pm$0.10 & 0/I & X-ray\\
MIR-37 & 21:40:39.80 &  58:18:24.9 &   \ldots  &     \ldots  & \ldots &     0.53$\pm$0.01&     0.76$\pm$0.01&     1.15$\pm$0.02&
1.39$\pm$0.03&  \ldots& 0/I \\
MIR-38 & 21:40:39.84 &  58:18:34.8 &   18.29$\pm$\ldots &  15.30$\pm$0.12 & 13.36$\pm$0.04  &      6.88$\pm$0.03& 8.19$\pm$0.03&
9.95$\pm$0.06&    13.49$\pm$0.06&     30.69$\pm$0.07 & II& X-ray \\
MIR-39 & 21:40:40.15 &  58:16:18.2 &   \ldots  &     \ldots  & \ldots &     1.36$\pm$0.01&     2.65$\pm$0.01&     3.24$\pm$0.04&
4.25$\pm$0.05&  \ldots & 0/I  & \\
MIR-40 & 21:40:40.43 &  58:18:26.1 &   18.28$\pm$\ldots & 15.64$\pm$0.15 & 13.85 $\pm$0.06 &     2.35$\pm$0.01&     2.05$\pm$0.01&
1.64$\pm$0.03&     0.76$\pm$0.02&  \ldots & III/S\\
MIR-41 & 21:40:40.96 &  58:16:12.0 &   17.55$\pm$\ldots & 15.89$\pm$\ldots  & 13.59$\pm$ 0.06  &    13.04$\pm$0.04& 23.41$\pm$0.06&
29.43$\pm$0.10&    42.25$\pm$0.15&  \ldots & 0/I \\
MIR-42 & 21:40:40.98 &  58:16:52.4 &    17.93$\pm$\ldots & 15.34$\pm$0.14 & 13.93$\pm$0.06 &     1.41$\pm$0.01&     1.04$\pm$0.01& 0.83$\pm$0.02&     0.37$\pm$0.03&  \ldots& III/S \\
MIR-43 & 21:40:41.07 &  58:13:58.9 &      12.96$\pm$0.02  &    12.08$\pm$ 0.03  &    11.77$\pm$ 0.02 &     6.80$\pm$0.02&     4.58$\pm$0.02&
3.19$\pm$0.03&     1.84$\pm$0.03&  \ldots& III/S  & \ha, X-ray\\
MIR-44 & 21:40:41.10 &  58:17:54.1 &      14.20$\pm$0.04  &    13.46$\pm$ 0.04  &    13.21$\pm$ 0.03 &     1.89$\pm$0.01&     1.14$\pm$0.01&
0.73$\pm$0.02&     0.37$\pm$0.02&  \ldots& III/S \\
MIR-45 & 21:40:41.14 &  58:15:11.2 &    12.97$\pm$0.03  &    11.61$\pm$ 0.04  &    10.68$\pm$ 0.02 &    66.38$\pm$0.17&    65.98$\pm$0.14&
81.84$\pm$0.21&   124.60$\pm$0.31&    194.40$\pm$0.09& II  & \ha, X-ray\\
MIR-46 & 21:40:41.27 &  58:17:14.2 &  \ldots  &     \ldots  &     \ldots &     0.06$\pm$0.00&     0.03$\pm$0.00&     0.65$\pm$0.02&
2.61$\pm$0.03&  \ldots & I/II\\
MIR-47 & 21:40:41.40 &  58:14:17.5 &   15.48$\pm$\ldots & 15.50$\pm$0.12 & 14.63$\pm$0.00 &     0.34$\pm$0.00&     0.21$\pm$0.00&
0.16$\pm$0.01&     0.10$\pm$0.02&  \ldots& III/S \\
MIR-48 & 21:40:41.43 &  58:16:37.8 &    18.30$\pm$\ldots & 15.60$\pm$0.15 &  12.79$\pm$0.03 &  51.81$\pm$0.14& 109.40$\pm$0.20&
158.50$\pm$0.34&   199.60$\pm$0.44& 1362.00$\pm$0.26 & 0/I\\
MIR-49 & 21:40:41.54 &  58:14:25.5 &    13.65$\pm$0.03  &    12.62$\pm$ 0.03  &    12.17$\pm$ 0.03 &     8.70$\pm$0.03&     7.38$\pm$0.03&
6.65$\pm$0.04&     6.42$\pm$0.04&  \ldots & II & \ha, X-ray\\
MIR-50 & 21:40:41.71 &  58:16:12.6 &     \ldots  &     \ldots  & \ldots &    10.84$\pm$0.04&    78.01$\pm$0.16&   128.00$\pm$0.29&
137.00$\pm$0.34&   1871.00$\pm$0.31 & 0/I & X-ray\\
MIR-51 & 21:40:41.89 &  58:15:23.0 &  15.68$\pm$0.09  &    14.30$\pm$ 0.06  &    13.65$\pm$ 0.05 &     2.88$\pm$0.02&     2.30$\pm$0.01&
3.75$\pm$0.04&     8.33$\pm$0.06&  \ldots & I/II & X-ray, {\em $JHK_s$}\\
MIR-52 & 21:40:41.96 &  58:13:49.1 &     15.50$\pm$0.06  &    15.00$\pm$ 0.09  &    14.98$\pm$ 0.14 &     0.40$\pm$0.00&     0.27$\pm$0.00&
0.19$\pm$0.01&     0.10$\pm$0.02&  \ldots & III/S\\
MIR-53 & 21:40:42.35 &  58:18:40.1 &   \ldots  &     \ldots  &     \ldots &     0.48$\pm$0.00&     2.16$\pm$0.01&     1.99$\pm$0.03&
1.26$\pm$0.02&  \ldots & H$_2$ \\
MIR-54 & 21:40:42.77 &  58:16:01.1 &    \ldots  &     \ldots  & \ldots &    20.16$\pm$0.06&    75.86$\pm$0.15&   113.50$\pm$0.27&
117.60$\pm$0.31&   2484.00$\pm$0.38 & I  & X-ray\\
MIR-55\tablenotemark{a} & 21:40:43.32 &  58:17:16.5 &    \ldots &     $>$17.6  & 15.31$\pm$0.11 &     1.65$\pm$0.01&     2.23$\pm$0.01&     2.72$\pm$0.03&
4.31$\pm$0.03&    145.40$\pm$0.08  & 0/I\\
MIR-56 & 21:40:43.34 &  58:18:42.6 &  16.32$\pm$0.13  &    14.42$\pm$ 0.05  &    13.55$\pm$ 0.04 &     1.93$\pm$0.01&     1.39$\pm$0.01&
0.91$\pm$0.02&     0.26$\pm$0.03&  \ldots & III/S\\
MIR-57 & 21:40:43.62 &  58:16:18.8 &   17.89$\pm$\ldots & 16.09$\pm$\ldots  & 13.51$\pm$ 0.07  &    33.21$\pm$0.09& 68.55$\pm$0.14&
82.72$\pm$0.21&    91.38$\pm$0.24&  \ldots & 0/I & X-ray \\
MIR-58 & 21:40:43.65 &  58:14:18.4 &   15.55$\pm$0.08  &    15.05$\pm$ 0.10  &    14.89$\pm$ 0.14 &     0.41$\pm$0.00&     0.28$\pm$0.00&
0.19$\pm$0.01&     0.14$\pm$0.02&  \ldots& Trans. \\
MIR-59 & 21:40:44.23 &  58:16:52.2 &  \ldots  &     \ldots  & \ldots &     1.36$\pm$0.01&     4.43$\pm$0.02&     9.13$\pm$0.05&
10.52$\pm$0.06&    100.90$\pm$0.07 & 0/I \\
MIR-60 & 21:40:44.26 &  58:14:12.4 &   15.57$\pm$0.06  &    15.00$\pm$ 0.09  &    14.69$\pm$ 0.12 &     0.45$\pm$0.00&     0.30$\pm$0.00&
0.22$\pm$0.01&     0.09$\pm$0.02&  \ldots& III/S \\
MIR-61 & 21:40:44.31 &  58:15:13.2 &   16.05$\pm$0.10  &    14.59$\pm$ 0.07  &    13.60$\pm$ 0.05 &     3.78$\pm$0.02&     4.10$\pm$0.02&
7.89$\pm$0.05&    15.74$\pm$0.08&  \ldots & I/II & X-ray, {\em $JHK_s$}\\
MIR-62 & 21:40:44.64 &  58:17:04.9 &   17.83$\pm$\ldots & 15.75$\pm$0.15  & 14.62$\pm$0.10 &     2.16$\pm$0.01& 2.38$\pm$0.01&
2.51$\pm$0.03&     3.06$\pm$0.03&  \ldots & II \\
MIR-63 & 21:40:44.82 &  58:16:05.0 &   15.86$\pm$\ldots & 14.28$\pm$0.06  & 12.89$\pm$0.05 &     7.08$\pm$0.03& 7.17$\pm$0.03&
6.83$\pm$0.05&     7.25$\pm$0.05&  \ldots & II  & X-ray, {\em $K_s$}\\
MIR-64 & 21:40:44.83 &  58:15:03.3 &    14.62$\pm$0.04  &    13.35$\pm$ 0.04  &    12.66$\pm$ 0.03 &     5.64$\pm$0.02&     5.23$\pm$0.02&
7.28$\pm$0.05&    12.94$\pm$0.06&  \ldots & I/II & \ha, X-ray\\
MIR-65 & 21:40:44.89 &  58:17:47.7 &   \ldots  &     \ldots  & \ldots &     0.42$\pm$0.00&     2.56$\pm$0.01&     2.01$\pm$0.02&
2.29$\pm$0.03&  \ldots & H$_2$\\
MIR-66\tablenotemark{b} & 21:40:45.11 &  58:15:24.0 &   18.46$\pm$0.02 &  16.74$\pm$0.01
& 15.71$\pm$0.01 &     0.59$\pm$0.01&     0.44$\pm$0.01&     1.73$\pm$0.03&
5.26$\pm$0.05&  \ldots & I/II & \\
MIR-67 & 21:40:45.16 &  58:15:59.6 &    \ldots  &     \ldots  & \ldots &     5.73$\pm$0.02&     9.26$\pm$0.03&    11.46$\pm$0.06&
11.78$\pm$0.07&  \ldots & I& X-ray \\
MIR-68\tablenotemark{b} & 21:40:45.24 &  58:16:41.3 &    17.38$\pm$0.01 &  16.29$\pm$ 0.01 &  15.53$\pm$0.01 &     0.43$\pm$0.01&     0.42$\pm$0.01&     0.41$\pm$0.02&
0.68$\pm$0.02&  \ldots & Galaxy\\
MIR-69 & 21:40:45.29 &  58:16:53.3 &   \ldots  &     \ldots  &     \ldots &     0.53$\pm$0.01&     0.61$\pm$0.01&     0.60$\pm$0.02&
\ldots&  \ldots & III/S\\
MIR-70\tablenotemark{b} & 21:40:45.41 &  58:16:45.5 &    20.38$\pm$0.15 &   17.48$\pm$0.02 &  15.96$\pm$0.01 &  0.33$\pm$0.00&
0.27$\pm$0.01&     0.48$\pm$0.02& 1.14$\pm$0.02&  \ldots & Galaxy\\
MIR-71 & 21:40:45.48 &  58:15:11.5 &    14.65$\pm$0.05  &    13.71$\pm$ 0.06  &    13.11$\pm$ 0.04 &     3.17$\pm$0.02&     2.61$\pm$0.01&
3.41$\pm$0.04&     5.73$\pm$0.05&  \ldots & I/II & X-ray, {\em $K_s$}\\
MIR-72 & 21:40:45.50 &  58:16:02.6 &    15.68$\pm$0.09  &    13.73$\pm$ 0.06  &    12.85$\pm$ 0.04 &     4.13$\pm$0.02&     3.15$\pm$0.02&
2.74$\pm$0.03&     2.88$\pm$0.03&  \ldots & II & X-ray\\
MIR-73 & 21:40:45.76 &  58:15:48.9 &  \ldots  &     \ldots  & \ldots &     2.71$\pm$0.02&     7.27$\pm$0.03&    10.79$\pm$0.06&
11.29$\pm$0.07&  \ldots & I & X-ray\\
MIR-74 & 21:40:45.82 &  58:17:43.0 &   \ldots  &     \ldots  & \ldots &     0.36$\pm$0.00&     0.92$\pm$0.01&     1.16$\pm$0.02&
1.38$\pm$0.02&  \ldots & H$_2$\\
MIR-75 & 21:40:46.22 &  58:17:40.2 &  \ldots  &     \ldots  &     \ldots &     0.34$\pm$0.00&     0.71$\pm$0.01&     1.20$\pm$0.02&
1.53$\pm$0.02&  \ldots & H$_2$\\
MIR-76 & 21:40:46.46 &  58:15:23.0 &    12.81$\pm$0.03  &    11.95$\pm$ 0.03  &    11.65$\pm$ 0.02 &     8.66$\pm$0.03&     5.35$\pm$0.02&
5.25$\pm$0.04&     7.74$\pm$0.07&  \ldots & Trans. & X-ray, \ha\\
MIR-77 & 21:40:46.86 &  58:15:33.4 &  15.30$\pm$0.06  &    13.55$\pm$ 0.05  &    12.63$\pm$ 0.04 &     5.80$\pm$0.02&     4.05$\pm$0.02&
5.16$\pm$0.04&    10.41$\pm$0.08&  \ldots & Galaxy & X-ray \\
MIR-78 & 21:40:47.35 &  58:16:37.3 &    17.08$\pm$\ldots & 16.14$\pm$0.24 & 14.63$\pm$0.10 &     1.00$\pm$0.01&     0.75$\pm$0.01&
0.59$\pm$0.02&     0.36$\pm$0.02&  \ldots & III/S\\
MIR-79 & 21:40:47.43 &  58:17:39.1 &    11.53$\pm$0.03  &     9.98$\pm$ 0.03  &     9.37$\pm$ 0.02 &    72.47$\pm$0.17&    46.36$\pm$0.10&
34.42$\pm$0.11&    20.21$\pm$0.07&  \ldots & III/S\\
MIR-80 & 21:40:48.00 &  58:15:37.6 &     13.89$\pm$0.03  &    12.95$\pm$ 0.03  &    12.67$\pm$ 0.03 &     4.72$\pm$0.02&     3.39$\pm$0.02&
5.77$\pm$0.04&    11.90$\pm$0.06&  \ldots & I/II & \ha, X-ray\\
MIR-81 & 21:40:48.52 &  58:16:37.1 &   16.18$\pm$0.09  &    14.95$\pm$ 0.08  &    14.46$\pm$ 0.08 &     0.71$\pm$0.01&     0.46$\pm$0.01&
0.38$\pm$0.02&     0.62$\pm$0.02&  \ldots & Galaxy& {\em $K_s$} \\
MIR-82 & 21:40:48.58 &  58:16:00.9 &    15.88$\pm$0.09  &    13.45$\pm$ 0.04  &    11.88$\pm$ 0.02 &    30.35$\pm$0.08&    37.01$\pm$0.08&
43.12$\pm$0.13&    61.94$\pm$0.17&    167.10$\pm$0.09 & II& {\em JH}\\
MIR-83 & 21:40:48.84 &  58:16:22.8 &   16.02$\pm$0.12  &    15.25$\pm$ 0.13  &    14.46$\pm$ 0.08 &     0.61$\pm$0.01&     0.43$\pm$0.01&
0.25$\pm$0.02& \ldots&  \ldots & III/S \\
MIR-84 & 21:40:48.87 &  58:15:10.9 &   15.64$\pm$0.06  &    15.04$\pm$ 0.08  &    14.56$\pm$ 0.09 &     0.77$\pm$0.01&     0.62$\pm$0.01&
0.75$\pm$0.02&     1.34$\pm$0.03&  \ldots & I/II & \ha, {\em $JK_s$}\\
MIR-85 & 21:40:49.05 &  58:17:19.4 &   15.79$\pm$0.08  &    13.83$\pm$ 0.05  &    12.92$\pm$ 0.02 &     3.32$\pm$0.02&     2.19$\pm$0.01&
1.75$\pm$0.02&     1.18$\pm$0.02&  \ldots & Trans. & {\em J} \\
MIR-86 & 21:40:49.06 &  58:17:09.3 &     14.14$\pm$0.03  &    12.86$\pm$ 0.04  &    12.13$\pm$ 0.02 &    13.88$\pm$0.05&    14.57$\pm$0.04&
15.54$\pm$0.07&    17.83$\pm$0.07&     17.99$\pm$0.05 & II & \ha, X-ray, {\em $K_s$}\\
MIR-87 & 21:40:49.29 &  58:14:11.3 &  13.52$\pm$0.03  &    12.66$\pm$ 0.03  &    12.35$\pm$ 0.02 &     3.87$\pm$0.02&     2.44$\pm$0.01&
1.66$\pm$0.02&     0.99$\pm$0.02&  \ldots & III/S\\
MIR-88 & 21:40:49.67 &  58:17:37.3 &   17.13$\pm$0.23  &    14.99$\pm$ 0.09  &    14.23$\pm$ 0.07 &     1.07$\pm$0.01&     0.73$\pm$0.01&
0.60$\pm$0.02&     0.36$\pm$0.02&  \ldots & III/S \\
MIR-89\tablenotemark{b} & 21:40:51.07 &  58:16:24.3 & 17.80$\pm$0.01 &  16.70$\pm$0.01 & 16.09$\pm$0.01 &   0.11$\pm$0.00&
0.08$\pm$0.00&     0.11$\pm$0.02& 0.23$\pm$0.02&  \ldots & Galaxy \\
MIR-90 & 21:40:51.23 &  58:16:38.9 &    14.23$\pm$0.03  &    13.76$\pm$ 0.04  &    13.61$\pm$ 0.04 &     1.26$\pm$0.01&     0.78$\pm$0.01&
0.57$\pm$0.02&     0.53$\pm$0.02&  \ldots & Trans. &{\em HK} \\
MIR-91 & 21:40:51.81 &  58:16:29.1 &   15.70$\pm$0.06  &    14.45$\pm$ 0.06  &    13.93$\pm$ 0.07 &     1.19$\pm$0.01&     0.74$\pm$0.01&
0.60$\pm$0.02&     0.54$\pm$0.02&  \ldots & Trans. &{\em HK} \\
MIR-92 & 21:40:51.90 &  58:17:01.6 &   \ldots  &     \ldots  &     \ldots &     0.36$\pm$0.00&     0.33$\pm$0.00&     0.20$\pm$0.02&
\ldots&  \ldots & III/S\\
MIR-93\tablenotemark{b} & 21:40:51.97 &  58:16:23.6 &  19.21$\pm$0.06 & 17.47$\pm$0.01 & 16.39$\pm$0.01 & 0.16$\pm$0.00& 0.10$\pm$0.00&     0.23$\pm$0.02& 0.65$\pm$0.02&  \ldots &Galaxy\\
MIR-94 & 21:40:52.26 &  58:17:46.0 &    15.68$\pm$0.05  &    14.56$\pm$ 0.05  &    14.06$\pm$ 0.06 &     0.94$\pm$0.01&     0.65$\pm$0.01&
0.59$\pm$0.02&     0.48$\pm$0.03&  \ldots & Trans. \\
MIR-95 & 21:40:52.98 &  58:18:01.1 &    15.29$\pm$0.05  &    14.59$\pm$ 0.06  &    14.22$\pm$ 0.07 &     0.84$\pm$0.01&     0.53$\pm$0.01&
0.54$\pm$0.03&     0.34$\pm$0.03&  \ldots &III/S \\
MIR-96 & 21:40:53.12 &  58:14:57.4 &  
    14.77$\pm$0.06  &    14.25$\pm$ 0.07  &    13.95$\pm$ 0.07 &     0.84$\pm$0.01&     0.54$\pm$0.01&
0.38$\pm$0.02&     0.37$\pm$0.03&  \ldots & Trans. \\
MIR-97 & 21:40:53.42 &  58:14:51.9 &     15.50$\pm$0.06  &    14.88$\pm$ 0.09  &    14.74$\pm$ 0.11 &     0.44$\pm$0.00&     0.31$\pm$0.01&
0.23$\pm$0.01&     0.20$\pm$0.02&  \ldots & Trans.\\
MIR-98 & 21:40:53.68 &  58:16:41.2 &   \ldots  &     \ldots  &     \ldots &     0.25$\pm$0.00&     0.71$\pm$0.01&     0.79$\pm$0.02&
0.83$\pm$0.02&  \ldots & H$_2$ \\
MIR-99 & 21:40:55.68 &  58:15:50.8 &   \ldots  &     \ldots  & \ldots &     0.36$\pm$0.00&     0.14$\pm$0.00&     2.69$\pm$0.03&
9.65$\pm$0.05&  \ldots & I/II \\
MIR-100 & 21:40:55.87 &  58:14:16.3 &     14.18$\pm$0.03  &    13.61$\pm$ 0.04  &    13.39$\pm$ 0.04 & \ldots&     0.87$\pm$0.02& \ldots&
0.45$\pm$0.03&  \ldots & III/S \\
MIR-101 & 21:40:55.88 &  58:15:34.8 &   \ldots  &     \ldots  & \ldots &     0.25$\pm$0.00&     0.32$\pm$0.00&     0.41$\pm$0.02&
0.71$\pm$0.02&  \ldots & Galaxy\\
MIR-102 & 21:40:55.93 &  58:17:07.8 &  14.45$\pm$0.03  &    13.19$\pm$ 0.04  &    12.76$\pm$ 0.03 &     3.21$\pm$0.02&     2.01$\pm$0.01&
1.39$\pm$0.02&     0.99$\pm$0.02&  \ldots & Trans. \\
MIR-103 & 21:40:57.04 &  58:16:56.8 &    16.61$\pm$0.13  &    15.92$\pm$ 0.21  &    16.22$\pm$ 0.00 &     0.42$\pm$0.01&     0.34$\pm$0.00&
0.40$\pm$0.02&     1.19$\pm$0.03&  \ldots & Galaxy \\
MIR-104 & 21:40:57.06 &  58:16:29.0 &  14.73$\pm$0.03  &    13.00$\pm$ 0.03  &    12.24$\pm$ 0.02 &     6.10$\pm$0.02&     4.11$\pm$0.02&
2.87$\pm$0.03&     1.96$\pm$0.03&  \ldots & Trans. \\
MIR-105 & 21:40:57.33 &  58:16:41.7 &   16.64$\pm$\ldots & 15.41$\pm$\ldots  & 14.55$\pm$ 0.13  &    0.79$\pm$0.01& 1.18$\pm$0.01&
2.90$\pm$0.03&     6.34$\pm$0.04&  \ldots & H$_2$\\
MIR-106 & 21:40:57.50 &  58:14:43.6 &     15.00$\pm$0.04  &    14.66$\pm$ 0.07  &    14.34$\pm$ 0.09 & \ldots&     0.43$\pm$0.01&
0.25$\pm$0.03&     0.15$\pm$0.02&  \ldots & III/S\\
MIR-107 & 21:40:58.03 &  58:15:12.1 &    15.06$\pm$0.05  &    14.50$\pm$ 0.07  &    14.12$\pm$ 0.08 &     0.70$\pm$0.01&     0.46$\pm$0.01&
0.30$\pm$0.02&     0.21$\pm$0.02&  \ldots & Trans. \\
MIR-108 & 21:40:58.45 &  58:16:20.5 &     13.11$\pm$0.03  &    12.44$\pm$ 0.03  &    12.02$\pm$ 0.03 &     5.93$\pm$0.03&     3.77$\pm$0.02&
3.13$\pm$0.03&     2.52$\pm$0.03&  \ldots & Trans. \\
MIR-109 & 21:40:58.53 &  58:15:00.3  &
   12.15$\pm$0.03  &    11.26$\pm$ 0.03  &    10.96$\pm$ 0.02 & \ldots&     8.34$\pm$0.03&
5.73$\pm$0.05&     3.46$\pm$0.03&  \ldots  & III/S\\
MIR-110 & 21:40:59.71 &  58:17:33.3  &   12.69$\pm$0.02  &    11.25$\pm$ 0.03  &    10.59$\pm$ 0.02 &    24.86$\pm$0.17& \ldots&
10.77$\pm$0.13& \ldots&  \ldots & III/S \\
\enddata
\tablenotetext{a}{$JHK_s$ photometry from \citet{nisini01}}
\tablenotetext{b}{$JHK_s$ photometry from \citet{beltran09}}
\tablenotetext{c}{Variable in $J/H/K_s$ }
\end{deluxetable}

\setcounter{table}{1}
\begin{deluxetable}{lccccccc}
\tabletypesize{\scriptsize}
\tablecaption{Coordinates and flux densities of NIR \& IRAC 3.6/4.5 $\mu$m
PMS sources in SFO~38
\label{tab_nirpms}}
\tablewidth{0pt}
\tablehead{
\colhead{Source} &
\colhead{$\alpha_{2000}$}&
\colhead{$\delta_{2000}$}&
\colhead{$J$}&
\colhead{$H$}&
\colhead{$K_{\rm s}$}&
\colhead{$F_{3.6}$}&
\colhead{$F_{4.5}$}\\
SFO38 & & & &  & &mJy &mJy \\}
\startdata
NIR-1 & 21:40:34.36 &  58:16:19.2 & 20.88$\pm$\ldots & 18.12$\pm$0.03 &
16.68$\pm$0.02 & 15.73$\pm$0.03 & 15.38$\pm$0.03\\
NIR-2 & 21:40:36.46 &  58:16:28.7 & 20.84$\pm$\ldots & 19.31$\pm$0.10 &
17.14$\pm$0.03 & 15.23$\pm$0.02 & 14.75$\pm$0.02\\
NIR-3 & 21:40:37.57 &  58:16:01.2 & 20.80$\pm$\ldots & 19.98$\pm$\ldots &
17.35$\pm$\ldots & 15.22$\pm$0.03 & 14.29$\pm$0.02\\
NIR-4 & 21:40:38.17 &  58:15:32.7 & 20.87$\pm$\ldots & 18.32$\pm$0.03 &
16.72$\pm$0.02 & 14.84$\pm$0.02 & 14.59$\pm$0.02\\
NIR-5 & 21:40:39.53 &  58:15:46.9 &  \ldots &  \ldots &   \ldots &
14.22$\pm$0.01 & 13.72$\pm$0.01\\
NIR-6 & 21:40:39.65 &  58:15:06.6 & 16.59$\pm$0.01 & 15.88$\pm$0.01 &
15.52$\pm$0.01 & 14.07$\pm$0.01 & 13.82$\pm$0.01\\
NIR-7 & 21:40:42.81 &  58:15:45.3 & 16.29$\pm$0.00 & 15.38$\pm$0.00 &
14.77$\pm$0.00 & 13.91$\pm$0.01 & 13.27$\pm$0.01\\
NIR-8 & 21:40:44.15 &  58:15:49.6 &  \ldots &  \ldots &   \ldots &
15.78$\pm$0.04 & 13.83$\pm$0.01\\
NIR-9 & 21:40:44.57 &  58:15:51.6 & 15.97$\pm$0.00 & 15.23$\pm$0.00 &
14.71$\pm$0.00 & 14.28$\pm$0.02 & 13.68$\pm$0.01\\
NIR-10 & 21:40:45.85 &  58:15:40.6 & 18.70$\pm$0.03 & 17.09$\pm$0.01 &
16.01$\pm$0.01 & 14.68$\pm$0.02 & 14.18$\pm$0.02\\
NIR-11 & 21:40:48.28 &  58:16:18.8 & 18.17$\pm$0.02 & 17.16$\pm$0.01 &
16.40$\pm$0.02 & 15.63$\pm$0.03 & 15.24$\pm$0.03\\
NIR-12 & 21:40:50.55 &  58:16:01.8 & 20.84$\pm$\ldots & 19.65$\pm$0.10 &
17.55$\pm$0.03 & 15.65$\pm$0.03 & 15.29$\pm$0.03\\
NIR-13 & 21:40:51.14 &  58:17:01.5 & 20.86$\pm$\ldots & 18.74$\pm$0.05 &
17.23$\pm$0.04 & 15.99$\pm$0.03 & 15.68$\pm$0.04\\
\enddata
\end{deluxetable}

\setcounter{table}{2}
\begin{deluxetable}{lccccc}
\tabletypesize{\scriptsize}
\tablecaption{{\em BVRI} photometry of MIR sources in SFO~38
\label{tab_optmag}}
\tablewidth{0pt}
\tablehead{
\colhead{Source} &
\colhead{$B$}&
\colhead{$V$}&
\colhead{$R$}&
\colhead{$I$}&
\colhead{Variability}\\}
\startdata
  MIR-1 &  19.82$\pm$0.010  & 18.24$\pm$0.005  &17.18$\pm$0.004  &16.00$\pm$0.003  &        \\
  MIR-2 &  20.48$\pm$0.015  & 18.88$\pm$0.006  &17.88$\pm$0.008  &16.84$\pm$0.005  &        \\
  MIR-3 &  21.58$\pm$0.021  & 19.58$\pm$0.007  &18.37$\pm$0.006  &17.14$\pm$0.004  &        \\
  MIR-4 &   0.00$\pm$0.000  & 20.38$\pm$0.016  &19.15$\pm$0.007  &17.73$\pm$0.006  &        \\
  MIR-5 &  21.11$\pm$0.015  & 19.32$\pm$0.007  &17.77$\pm$0.004  &16.37$\pm$0.002  & {\em BVI}       \\
  MIR-7 &  21.59$\pm$0.021  & 19.39$\pm$0.006  &17.90$\pm$0.004  &16.38$\pm$0.002  &        \\
  MIR-8 &  17.74$\pm$0.001  & 15.74$\pm$0.001  &14.55$\pm$0.001  &13.35$\pm$0.001  &        \\
  MIR-9 &  19.95$\pm$0.017  & 18.48$\pm$0.004  &17.60$\pm$0.005  &16.62$\pm$0.004  &        \\
 MIR-10 &  17.28$\pm$0.001  & 16.07$\pm$0.001  &15.29$\pm$0.001  &14.49$\pm$0.001  &        \\
 MIR-11 &   0.00$\pm$0.000  & 19.52$\pm$0.011  &17.94$\pm$0.007  &16.19$\pm$0.005  &        \\
 MIR-14 &  17.67$\pm$0.001  & 16.24$\pm$0.001  &15.38$\pm$0.001  &14.53$\pm$0.001  &        \\
 MIR-15 &  16.89$\pm$0.001  & 15.26$\pm$0.001  &14.33$\pm$0.001  &13.44$\pm$0.001  &        \\
 MIR-18 &  19.62$\pm$0.006  & 17.58$\pm$0.003  &16.30$\pm$0.002  &14.98$\pm$0.009  &        \\
 MIR-26 &  17.34$\pm$0.001  & 16.03$\pm$0.001  &15.20$\pm$0.001  &14.30$\pm$0.001  &        \\
 MIR-29 &  20.58$\pm$0.011  & 18.43$\pm$0.005  &16.94$\pm$0.002  &15.42$\pm$0.002  &        \\
 MIR-31 &  17.43$\pm$0.002  & 15.71$\pm$0.001  &14.60$\pm$0.001  &13.52$\pm$0.001  &        \\
 MIR-32 &  21.15$\pm$0.015  & 19.29$\pm$0.006  &18.01$\pm$0.007  &16.25$\pm$0.002  & {\em BVI}\\
 MIR-43 &  19.41$\pm$0.004  & 17.49$\pm$0.002  &16.21$\pm$0.001  &14.73$\pm$0.001  & {\em BVI}\\
 MIR-44 &  19.36$\pm$0.006  & 17.65$\pm$0.003  &16.57$\pm$0.001  &15.51$\pm$0.002  &        \\
 MIR-45 &  21.41$\pm$0.018  & 19.45$\pm$0.007  &18.10$\pm$0.012  &16.33$\pm$0.002  & {\em BVI}\\
 MIR-47 &   0.00$\pm$0.000  & 19.86$\pm$0.009  &18.76$\pm$0.005  &17.61$\pm$0.005  &        \\
 MIR-49 &  20.91$\pm$0.012  & 18.81$\pm$0.004  &17.04$\pm$0.003  &15.69$\pm$0.001  & {\em BVI}\\
 MIR-52 &  19.54$\pm$0.008  & 18.32$\pm$0.005  &17.52$\pm$0.003  &16.64$\pm$0.005  &        \\
 MIR-58 &  20.13$\pm$0.007  & 18.75$\pm$0.003  &17.86$\pm$0.005  &16.91$\pm$0.003  & {\em B}  \\
 MIR-60 &  20.79$\pm$0.011  & 19.13$\pm$0.005  &18.07$\pm$0.005  &17.06$\pm$0.003  &        \\
 MIR-64 &   0.00$\pm$0.000  & 20.49$\pm$0.014  &18.77$\pm$0.016  &17.09$\pm$0.006  &        \\
 MIR-76 &  19.08$\pm$0.006  & 17.16$\pm$0.002  &15.90$\pm$0.002  &14.55$\pm$0.001  &        \\
 MIR-79 &  21.41$\pm$0.024  & 18.56$\pm$0.004  &16.52$\pm$0.002  &14.55$\pm$0.001  &        \\
 MIR-80 &  21.69$\pm$0.023  & 19.81$\pm$0.008  &17.90$\pm$0.006  &16.00$\pm$0.004  &{\em V}   \\
 MIR-87 &  19.72$\pm$0.005  & 17.72$\pm$0.002  &16.50$\pm$0.002  &15.28$\pm$0.002  &        \\
 MIR-90 &  18.15$\pm$0.002  & 16.85$\pm$0.002  &16.04$\pm$0.001  &15.26$\pm$0.002  &        \\
 MIR-95 &   0.00$\pm$0.000  & 20.16$\pm$0.012  &18.71$\pm$0.009  &17.34$\pm$0.004  &        \\
 MIR-96 &  19.23$\pm$0.008  & 17.85$\pm$0.003  &16.94$\pm$0.003  &16.02$\pm$0.002  &        \\
 MIR-97 &  20.06$\pm$0.007  & 18.63$\pm$0.005  &17.76$\pm$0.005  &16.83$\pm$0.004  &        \\
MIR-100 &  18.31$\pm$0.003  & 16.86$\pm$0.002  &16.03$\pm$0.001  &15.28$\pm$0.002  &        \\
MIR-106 &  18.87$\pm$0.004  & 17.75$\pm$0.003  &16.96$\pm$0.003  &16.21$\pm$0.002  &        \\
MIR-107 &  19.50$\pm$0.004  & 18.12$\pm$0.002  &17.22$\pm$0.003  &16.27$\pm$0.002  &        \\
MIR-108 &  18.94$\pm$0.005  & 17.31$\pm$0.004  &16.13$\pm$0.003  &14.86$\pm$0.001  &        \\
MIR-109 &  18.17$\pm$0.003  & 16.17$\pm$0.001  &15.00$\pm$0.001  &13.86$\pm$0.001  &        \\
MIR-110 &   0.00$\pm$0.000  & 19.78$\pm$0.010  &17.76$\pm$0.004  &15.76$\pm$0.003  &{\em V}   \\
BRC38 1 &   0.00$\pm$0.000  & 20.05$\pm$0.017  &18.24$\pm$0.009  &16.45$\pm$0.003  &        \\
BRC38 17 &   0.00$\pm$0.000  & 20.51$\pm$0.017  &18.65$\pm$0.005  &16.92$\pm$0.006  & {\em V} \\
GFG 64 &  19.01$\pm$0.007  & 17.48$\pm$0.003  &16.19$\pm$0.001  &15.18$\pm$0.001  &        \\
IPHAS &   0.00$\pm$0.000  & 20.28$\pm$0.011  &18.28$\pm$0.009  &16.50$\pm$0.004  &        \\
\enddata
\end{deluxetable}

\setcounter{table}{3}
\begin{deluxetable}{llllllllllllc}
\tabletypesize{\scriptsize}
\rotate
\tablecaption{Spectral Classification of YSOs
\label{tab_sptclass}}
\tablewidth{0pt}
\tablehead{
\colhead{ID}&
\colhead{Alt. ID} &
\colhead{Date}&
\colhead{W$_\lambda$}&
\colhead{Sp. Type}&
\colhead{T$_{eff}$} &
\colhead{Type}&
\colhead{\ha[10$\%$]}&
\colhead{log \.{M}$_{ac}$}&
\colhead{$\alpha_{IRAC}$} &
\colhead{E(V-I)} &
\colhead{A$_{V}$}&
\colhead{E(V-I)/E(B-V)}\\
        &           &          & (\AA)     &($\pm$subtype)& (K)&     & (km s$^{-1})$  &(\msun yr$^{-1}$)    &          &       &        &      \\}
\startdata
BRC 38 1&           &   5/10/08&  -12.93 & M3$\pm$1&  3470&  CTTS&  312.23 &  -9.86   &         &  1.13 &   2.81 &       \\ 
        &           &  10/10/09&  -16.24 &         &      &      &  205.29 &  -10.90  &         &       &        &       \\            
MIR-5   &  BRC 38 2 &   5/10/08&  -85.23 & M1$\pm$1&  3720&  CTTS&  359.79 &  -9.40   &         &  0.99 &   2.47 &  3.19 \\ 
        &           &  10/10/09&  -106.1 &         &      &      &  244.53 &  -10.52  &         &       &        &       \\            
MIR-7   &  BRC 38 3 &   6/10/08&  -19.0  & M1$\pm$2&  3720&  CTTS&  476.25 &  -8.27   & -0.87   &  1.05 &   2.61 &  1.46 \\ 
        &           &  10/10/09&  -18.75 &         &      &      &  463.50 &  -8.39   &         &       &        &       \\            
MIR-11  &  BRC 38 4 &  10/09/08&  -7.30  & M1$\pm$2&  3720&  WTTS&         &          &  0.02   &  1.37 &   3.41 &       \\ 
        &           &  18/07/09&  -11.29 &         &      &      &         &          &         &       &        &       \\            
MIR-29  &  BRC 38 5 &   6/09/08&  -3.42  & M1$\pm$2&  3720&  WTTS&         &          & -1.74   &  1.05 &   2.61 &  1.57 \\ 
        &           &   3/11/08&  -8.84  &         &      &	 &  469.90 &  -8.33   &         &       &        &       \\
        &           &  24/08/09&  -1.89  &         &      &      & 	   &          &         &       &        &       \\	        
        &           &   9/11/09&  -2.74  &         &      &      &         &          &         &       &        &       \\  	      
MIR-31  &  BRC 38 6 &   6/09/08&  -53.47 & K2$\pm$1&  4900&  CTTS&  384.43 &  -9.16   & -1.60   &  1.18 &   2.94 &  1.42 \\ 
	&           &  24/08/09&  -57.32 &         &      &      &  442.02 &  -8.60   &         &       &        &       \\            
	&           &   9/10/09&  -60.5  &         &      &      &  382.26 &  -9.18   &         &       &        &       \\            
MIR-32  &  BRC 38 7 &  10/09/08&  -33.12 & M2$\pm$1&  3580&  CTTS&  368.03 &  -9.32   & -1.03   &  0.90 &   2.24 &  2.65 \\ 
	&	    & 12/09/09&  -15.78  &         &      &      &  220.59 &  -10.75  &         & 	&  	 &       \\
MIR-43  &  GFG 62   &  18/07/09&  -3.516 & M1$\pm$1&  3720&  WTTS&         &          & -2.61   &  0.80 &   1.99 &  1.82 \\ 
MIR-45  &  BRC 38 9 &   6/09/08&  -37.43 & K4$\pm$1&  4590&  CTTS&  281.09 &  -10.16  & -0.18   &  1.97 &   4.91 &  2.21 \\ 
 	&           &   3/11/08&  -41.22 &         &      &      &  389.74 &  -9.11   &         &       &        &       \\            
 	&           &  24/08/09&  -56.33 &         &      &      &  273.39 &  -10.24  &         & 	&  	 &       \\
MIR-49  &  BRC 38 10&   4/11/08&  -13.5  & M0$\pm$1&  3850&  CTTS&  430.70 &  -8.71   & -1.37   &  1.32 &   3.29 &  1.91 \\       
	&	    &  9/10/09&  -30     &         &      &      &  450.55 &  -8.52   & 	& 	&  	 &       \\  
MIR-64	&  BRC 38 11&  10/09/08&  -31.46 & K5$\pm$1&  4350&  CTTS&  239.20 &  -10.57  &  0.10   &  2.04 &   5.08 &       \\ 
 	&	    & 12/09/09&  -25.06  &         &      &      &         &          &  	&       &        &       \\            
MIR-76  &  GFG 81   &  18/07/09&  -3.261 & K7$\pm$1&  4060&  WTTS&         &          & -1.07   &  1.01 &   2.51 &  1.87 \\ 
	&	    & 10/10/09&  -3.135  &         &      &      &         &          &  	&       &        &       \\            
MIR-80  &  BRC 38 12&  24/08/09&  -8.404 & M4$\pm$1&  3370&  WTTS&         &          &  0.31   &  0.95 &   2.37 &  3.39 \\ 
BRC 38 17&          &  11/11/09&  -119.  & M3$\pm$2&  3470&  CTTS&  205.29 &  -10.90  &         &  1.12 &   2.79 &       \\
\enddata                    
\end{deluxetable}
\setcounter{table}{4}
\begin{deluxetable}{cccccccccc}
\tabletypesize{\small}
\tablecolumns{9}
\tablewidth{0pt}
\tablecaption{Parameters derived from SED modeling using axisymmetric
radiation transfer models for the candidate YSOs.  \av\ refers to
the foreground extinction towards the source. 
\label{tab_sedfits}}
\tablehead{
\colhead{Source} & \colhead{$T_{\ast}$} & \colhead{\mstar} & \colhead{$M_{\rm disk}$} & \colhead{$M_{\rm env}$} &
\colhead{$\dot{M}_{\rm env}$} & \colhead{$\dot{M}_{\rm disk}$} &
\colhead{$L$} & \colhead{\av} & \colhead{Age}\\
\colhead{}& \colhead{K} & \colhead{\msun} & \colhead{\rm M$_{\odot}$} & \colhead{\rm M$_{\odot}$} & \colhead{\rm M$_{\odot}$~yr$^{-1}$} &\colhead{\rm M$_{\odot}$~yr$^{-1}$} &
\colhead{\rm L$_{\odot}$} & & Myr
}
\startdata
MIR-11&  4442 &  1.26  & 6.44$\times10^{-3}$ &  6.39$\times10^{-5}$  & 0.0                 & 2.10$\times10^{-9}$ &1.27 &  5.3 & 3.2\\
MIR-29&  4837 &  1.55  & 1.66$\times10^{-3}$ &  6.13$\times10^{-7}$  & 0.0                 & 1.16$\times10^{-9}$ &1.67   &  4.4 & 5.8\\
MIR-31&  5923 &  1.88  & 1.33$\times10^{-4}$ &  7.04$\times10^{-8}$  & 0.0                 & 2.41$\times10^{-9}$ &8.92   &  4.2 & 6.7\\
MIR-34&  4523 &  1.53  & 7.64$\times10^{-5}$ &  2.56$\times10^{-3}$  & 1.21$\times10^{-8}$ & 1.28$\times10^{-10}$& 5.8 &  3.2  &0.69\\
MIR-36&  5413 &  3.29  & 4.37$\times10^{-3}$ &  1.85$\times10^{-8}$  & 0.0                 & 4.80$\times10^{-8}$ &29.1 &  49.6 & 1.3\\
MIR-38&  7854 &  1.98  & 3.10$\times10^{-5}$ &  2.99$\times10^{-6}$  & 0.0                 & 5.49$\times10^{-11}$&11.9  &  23.0 & 8.2\\
MIR-45&  9960 &  2.32  & 4.23$\times10^{-5}$ &  1.70$\times10^{-8}$  & 0.0                 & 1.37$\times10^{-9}$ &27.0 &  8.3 & 9.4\\
MIR-48&  7554 &  5.14  & 1.71$\times10^{-3}$ &  21.4$\times10^{-2}$  & 2.23$\times10^{-7}$ & 2.38$\times10^{-9}$ & 280 &45.9 & 0.49\\
MIR-50&  4643 &  5.97  & 7.67$\times10^{-2}$ &  40.8                 & 3.82$\times10^{-4}$ & 2.08$\times10^{-8}$ &196.7   &  44.4 &0.1\\
MIR-54&  4290 &  1.50  & 2.46$\times10^{-2}$ &  9.3                  & 7.13$\times10^{-5}$ & 2.98$\times10^{-6}$ &33.4  &  19.6 &0.17\\
MIR-55&  4989 &  3.07  & 7.86$\times10^{-4}$ &  1.44$\times10^{-3}$ & 1.97$\times10^{-9}$ & 1.14$\times10^{-9}$ &14.4   &  52.3 &0.91\\
MIR-59&  5218 &  4.27  & 9.99$\times10^{-3}$ &  7.37$\times10^{-3}$ & 2.95$\times10^{-7}$ & 2.00$\times10^{-8}$ & 60.5  &  87.4 & 0.53\\
MIR-82&  11320&  2.77  & 5.85$\times10^{-5}$ &  1.18$\times10^{-6}$  & 0.0                 & 1.40$\times10^{-10}$ &57.2&  21.0 & 4.8\\
MIR-86&  4594 &  1.55  & 8.18$\times10^{-4}$ &  3.63$\times10^{-7}$  & 0.0                 & 4.50$\times10^{-10}$ &2.59  &  9.7 & 1.8\\
\enddata
\end{deluxetable}

\end{document}